\documentclass[aps,prl,twocolumn,showpacs,preprintnumbers,amsmath,amssymb,
floatfix,superscriptaddress]{revtex4-1}

\newcommand{\hh}[1]{{\color{black}{#1}}}
\newcommand{\rere}[1]{{\color{black}{#1}}}

\usepackage{ulem}

\usepackage{graphics}
\usepackage{graphicx}
\usepackage{amsmath}
\usepackage{setspace}
\usepackage{braket}
\usepackage{epstopdf}
\usepackage{comment}
\usepackage{hyperref}
\usepackage{bm}
\usepackage{xfrac}
\usepackage{blindtext}
\usepackage{mathrsfs}
\usepackage{xcolor}

\newcommand{\affA}{Van der Waals-Zeeman Institute, Institute of Physics, University of Amsterdam, 1098 XH Amsterdam, Netherlands}
\newcommand{\affB}{QuSoft, Science Park 123, 1098 XG Amsterdam, the Netherlands}
\newcommand{\affC}{Institute for Theoretical Physics, Institute of Physics, University of Amsterdam, Science Park 904, 1098 XH Amsterdam, the Netherlands}
\newcommand{\affE}{Institute for Molecules and Materials, Radboud University, Heyendaalseweg 135, 6525 AJ Nijmegen, Netherlands}
\newcommand{\affD}{Fritz-Haber-Institut der Max-Planck-Gesellschaft, Faradayweg 4-6, 14195 Berlin, Germany }
\newcommand{\affF}{Department of Physics \rere{and Astronomy}, Stony Brook University, Stony Brook, New York 11794, USA}
\begin{document}

\title{Observation of Chemical Reactions between a Trapped Ion and Ultracold Feshbach Dimers}

\author{H.~Hirzler}\affiliation{\affA}
\author{R.\,S.~Lous}\affiliation{\affA}
\author{E.~Trimby}\affiliation{\affA}
\author{J.~P\'erez-R\'{i}os}\affiliation{\affD}\affiliation{\affE}\affiliation{\affF}
\author{A.~Safavi-Naini}\affiliation{\affB}\affiliation{\affC}
\author{R.~Gerritsma}\affiliation{\affA}\affiliation{\affB}

\date{\today}

\begin{abstract}

We measure chemical reactions between a single trapped $^{174}$Yb$^+$ ion \hh{and ultracold Li$_2$ dimers.} 
This produces LiYb$^+$ molecular ions that we detect via mass spectrometry. We explain the reaction rates by modelling the dimer density as a function of the magnetic field and obtain excellent agreement when we assume the reaction to follow the Langevin rate. Our results present a novel approach towards the creation of cold molecular ions and point to the exploration of ultracold chemistry in ion molecule collisions. What is more, with a detection sensitivity below molecule densities of $10^{14}\,\mathrm{m}^{-3}$,  we provide a new method to detect low-density molecular gases.
\end{abstract}

\maketitle


\paragraph{Introduction.}
To identify how quantum effects contribute to physical and chemical processes, it is essential to study chemical reactions at very low temperatures where only few partial waves play a role. Ion-molecule mixtures present a versatile platform to measure reaction channels with increased richness as compared to atomic mixtures. Interacting ions and molecules have been studied by letting ions collide with molecules from the vacuum background~\cite{Sugiyama_1995,Sugiyama1997,Hoang2020}, from inlet  room-temperature sources~\cite{Rugango2015} and from molecular beams~\cite{Willitsch:2008,JPRBook,Heazlewood2021}. In these studies molecule temperatures were in the $1\,$K range, which is far above the ion-molecule s-wave collision energies. However, molecular samples of \rere{much} lower temperatures can be created from ultracold atoms using Feshbach resonances~\cite{Koehler2006, Ferlaino:2009, Julienne:2010}. These weakly bound diatomic molecules are called Feshbach dimers. Merging the fields of trapped ions and ultra-cold quantum gases~\cite{Tomza:2019} paves the way for studying ion-molecule collisions in the ultracold regime. 

An important premise to control ion-neutral interactions in the ultracold regime is the understanding of the relevant reaction channels. {Examples are} charge transfer~\cite{Ratschbacher:2012, Haze:2015, Jyothi2016, Joger:2017}, spin exchange~\cite{Ratschbacher:2013, Sikorsky:2018, Fuerst:2018:spin} and three-body recombination \cite{Haerter2012,PerezRios2015,Kruekow2016,Mohammadi2021}. The exceptional control over the quantum states of trapped ions~\cite{Monroe:1995, Leibfried:2003} makes it possible to study these chemical reactions at the single particle level and gives direct experimental access to the reaction products, their quantum states and  energies, as well as their branching ratios. {Single trapped ions can thus be used as probes}  to detect properties of the ultracold gases in which they are immersed~\cite{Schmid:2010,Joger:2017}\rere{e.g. BEC-BCS crossover regime~\cite{Zwerger2012tbb} and the charged polarons}
%
~\cite{Tomza:2019, Astrakharchik:2021, Christensen2021cpa, Oghittu2021a}. Moreover, ultracold molecule-ion mixtures can be used to form cold molecular ions with applications in quantum information and precision spectroscopy~\cite{MurPetit:2012, Khanyile2015, Wolf:2016, Chou:2017, Sinhal:2020, Katz2021}

In this Letter, we report on the observation of {cold} collisions between single Yb$^+$ ions in a Paul trap and a mixture of ultracold Li atoms and Li$_2$ dimers. We study the occurence of chemical reactions
 by observing the Yb$^+$ fluorescence after it has interacted with the cloud, counting the number of times the Yb$^+$ ion goes dark. We measure a negative correlation between the dark events probability and the atom density, indicating that atoms are not involved. Instead, we find that the reaction $\mathrm{Li_2}+\mathrm{Yb^+}\rightarrow\mathrm{LiYb^+}+\mathrm{Li}$ leads to the dark events.  We use mass-spectrometry to demonstrate the occurrence of LiYb$^+$ molecular ions. We show excellent agreement between the probability of dark events and the Li$_2$ density in our system, which we model with rate equations. \rere{We use our ion sensor to detect about 50 dimers in a cloud of $\sim10^4$ atoms, which provides a new tool to detect molecules in sparse quantities. This creates a pathway to create cold molecular ions and to search for quantum effects in ion-molecule collisions. } 
 \rere{Using ultracold Feshbach dimers, our molecules are 2-5 orders of magnitude colder compared to previous ion-molecule studies ~\cite{Sugiyama_1995,Sugiyama1997,Hoang2020,Rugango2015,Willitsch:2008,JPRBook,Heazlewood2021}.}

\begin{figure}
	\centering
	\includegraphics[width=1\linewidth]{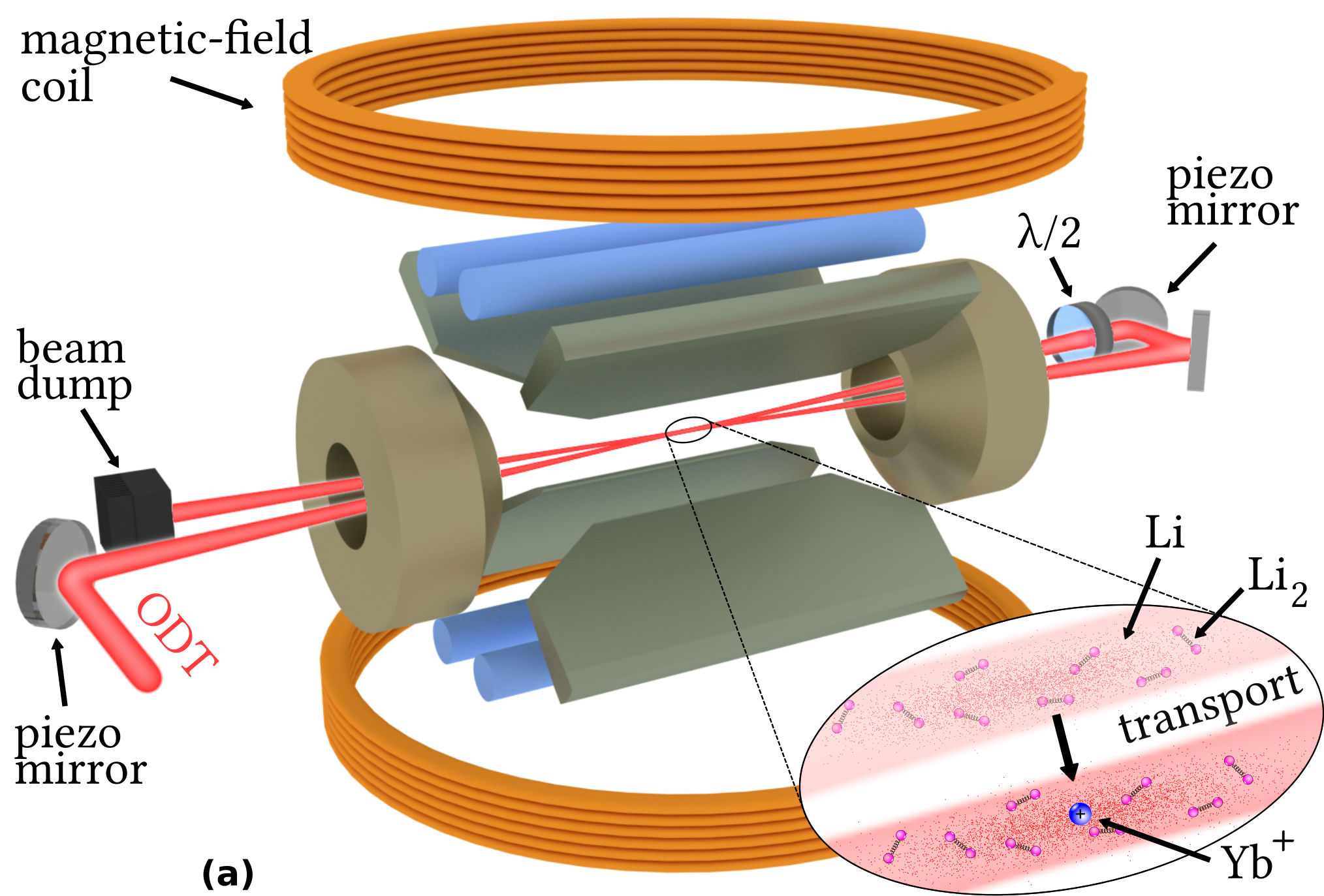}	
	\includegraphics[width=1\linewidth]{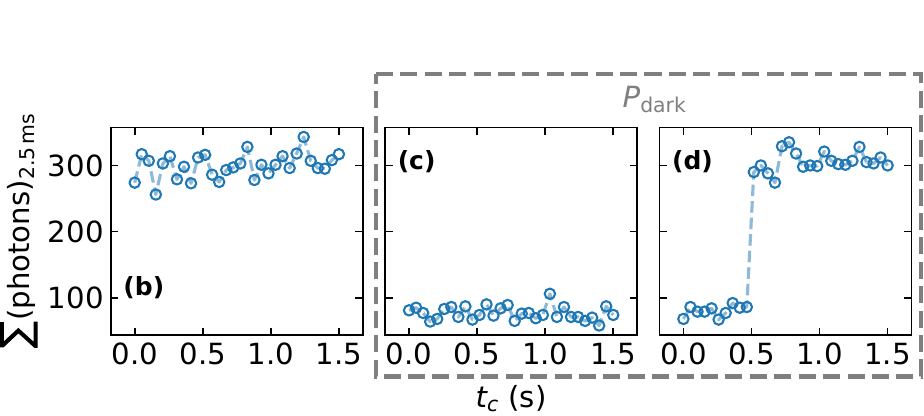}
 	\caption{Experimental set-up with optical dipole trap (ODT) and Paul trap (shown in grey) to confine atoms and single ions respectively. The lower panel shows single-run Yb$^+$ ion $2.5\,$ms fluorescence detection versus time after ion-bath interaction resulting in b) a bright Yb$^+$, c) a lost Yb$^+$ and d) an Yb$^+$ that turns bright after $t_\mathrm{c}\approx0.5\,$s of Doppler cooling.}
 	\label{fig:setup}
\end{figure}

\paragraph{Experimental sequence.}
The Yb$^+$-Li mixture is prepared in a hybrid ion-neutral trap as depicted in Fig.~\ref{fig:setup} and more extensively described in Ref.~\cite{Hirzler2020}. We load a single $^{174}$Yb$^+$ ion by isotope-selective two-photon ionization, Doppler-cool it to about 0.5\,mK and prepare it in the $^2S_{1/2}$ ground state. The ion trap operates at a driving frequency $\Omega=2\pi\times1.85\,$MHz and trap frequencies $(\omega_x, \omega_y, \omega_z) \approx\,2\pi\times(191, 196, 112)\,$kHz, where $z$ is the direction along the axis of the Paul trap. The ultracold fermionic Li atoms are prepared in a crossed $1070\,$nm, $40\,\mu$m waist, optical dipole trap (ODT)  about $200\,\mu$m below the ion, using forced evaporative cooling at 663\,G close to the 832\,G Feshbach resonance~\cite{Zuern:2013}. We obtain about $2.2\times10^4$ $^6$Li atoms per spin state in the lowest two magnetic sublevels $\ket{F=1/2, m_F=\pm 1/2}$  at a temperature $T=1-10\,\mu$K. Here $F$ is the total angular momentum quantum number and $m_F$ is its projection on the quantization axis.

We admix a small quantity of Li$_2$ dimers to the bath by setting the magnetic field to $B_\mathrm{Li_2}=693\,$G 
and associating the dimers by three-body recombination through direct evaporation. We do this in the final stage of the evaporation of the lithium spin mixture.  The number of resulting dimers depends primarily on the temperature, the magnetic field $B_\mathrm{Li_2}$, and the atom density $n_\mathrm{a}$~\cite{Jochim:2003, Chin2004}, {which influences the three-body recombination and dissociation rate of the $\mathrm{Li}+\mathrm{Li'}+\mathrm{Li}\rightleftharpoons\mathrm{Li_2}+\mathrm{Li}$ reactions.} Here, Li and Li$'$ indicate the two  spin states.
Next, we turn off the magnetic field, which increases the dimer binding energy to a fixed value of about $E_\mathrm{b}/h=1.38\,$GHz~\cite{Suppl:2021, Julienne:2010}, with $h$ being Planck's constant. By ramping to zero field we minimize variations in the molecular ion formation rate~\cite{Hirzler2020a}, as well as quantum effects~\cite{Tomza:2019, Weckesser:2021}. \rere{Moreover, as the binding energy ($\approx 70\,$mK) is much greater than the dimer-ion collision energy, molecular ion formation is expected to be the dominant reaction channel~\cite{Hirzler2020a}.} For all the reported experiments we {obtain} dimer densities that are less than 10\% of the atomic density. {We measure the atom observables by time-of-flight absorption imaging.}

We overlap the Yb$^+$ ion with the atom-dimer bath by transporting the ODT to the location of the Yb$^+$  ion by means of piezo-electric mirrors and let the systems interact for $\tau=500$~ms. \hh{The interaction time was chosen
to have sufficient contrast, yet reasonable experimental cycle time.} Subsequently, we Doppler cool the Yb$^+$ ion for $1500\,$ms and simultaneously use a photomultiplier tube (PMT) to detect its fluorescence in time bins of $50\,$ms.

From the PMT measurements after the interaction, we identify three possible outcomes for each experimental run, as is indicated in Fig.~\ref{fig:setup} (b-d). After the interaction the Yb$^+$ ion is either bright (panel b) or dark (panel c) indicating ion loss.  The final scenario occurs when the ion is initially dark, but after some cooling time $t_\mathrm{c}$, it turns  bright (panel d). 
 We count events c) and d) together as the dark ion probability $P_\mathrm{dark}$.
 
 \hh{The dark events hint towards the possibility of molecular ions being formed, 
 as they resemble what was recently seen in the Ba$^+$-Rb system~\cite{Mohammadi2021}. There, the photo-dissociation of the BaRb$^+$ molecular ion with light at $1064\,$nm, resulted in the observation of dark Ba$^+$ ions with similar fluorescence characteristics as panel d).} \hh{We will show that molecular ions in our system originate from ion-dimer collisions, via $\mathrm{Li_2}+\mathrm{Yb^+}\rightarrow\mathrm{LiYb^+}+\mathrm{Li}$ as it was proposed in~\cite{Hirzler2020a} in contrast to three-body recombination of molecular ions reported in Ref.~\cite{Mohammadi2021}.}



\begin{figure}
	\centering
	\includegraphics[width=\linewidth]{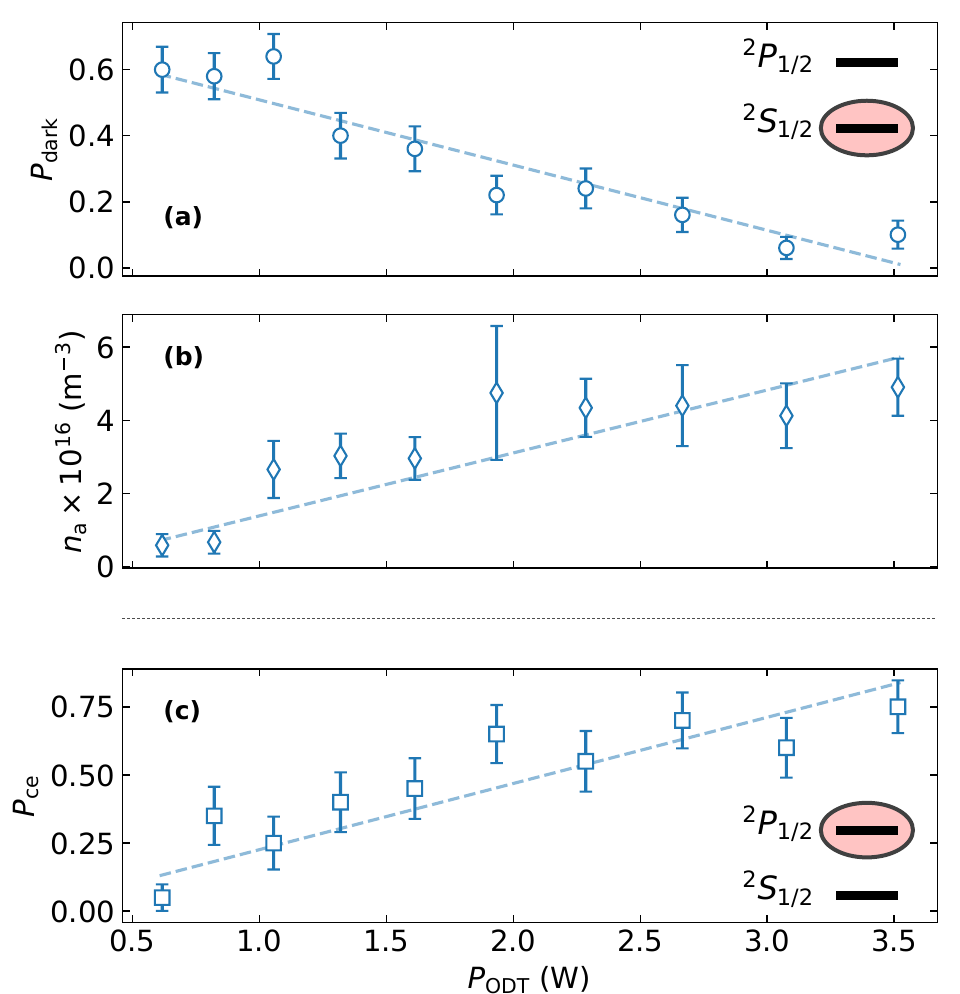}
 	\caption{Dark Yb$^+$ ion probability a) as a function of the ODT power with Yb$^+$ initialized in $^2S_{1/2}$. Peak atom density $n_\mathrm{a}$ b), obtained with time-of-flight measurements. Charge exchange probability c) for Yb$^+$ in its $^2P_{1/2}$ state. The dashed lines are linear fits to the data as a guide to the eye. The error bars reflect the total
statistical uncertainties.}
 	\label{fig:evapScan}
\end{figure}

\paragraph{Results. } We study \hh{the ion interacting with the atom-dimer bath and find a negative correlation between the probability of dark ions and the atom density}
, as shown in Fig.~\ref{fig:evapScan}. In a) we plot $P_\mathrm{dark}$ as a function of the ODT laser power $P_\mathrm{ODT}$ at the end of the evaporation ramp. The latter changes the peak atom density ($5-0.5\times10^{16}\,$m$^{-3}$), presented in b), as well as the temperature of the atom cloud ($12-0.5\,\mu$K) \hh{and the dimer density}~\cite{Suppl:2021}.

Additional charge exchange measurements of the process Yb$^+$($^2P_{1/2}$)+Li\,$\rightarrow$Li$^+$+Yb are taken as an independent means to probe the atomic density (see Fig.~\ref{fig:evapScan} c). Charge exchange occurs when we laser excite the ion to the $^2P_{1/2}$ state during the interaction with the bath and it directly leads to ion loss~\cite{Joger:2017}. To avoid signal saturation we use $\tau=50\,$ms. Since the charge exchange rate is independent of the collision energy, its rate is a direct probe of the local atom density around the ion. The measurements confirm the trend observed in Fig.~\ref{fig:evapScan} b.

\hh{The negative correlation between $P_\mathrm{dark}$ and $n_\mathrm{a}$, excludes Li atoms as the origin for the reaction  resulting in $P_\mathrm{dark}$ and points towards a role of Li$_2$. In particular, it excludes three-body recombination of molecular ions via $\mathrm{Li}+\mathrm{Li}+\mathrm{Yb^+}\rightarrow\mathrm{LiYb^+}+\mathrm{Li}$.}
This is supported by theory since the three-body recombination rate is given by $\Gamma_3=k_3 n_\mathrm{a}^2$, with coefficient $k_3 = 8\pi^2/15\sqrt{2/\mu}(2\alpha)^{5/4}E_\mathrm{col}^{-3/4}$, where $\alpha$ is the atomic polarizability, $\mu$ is the atom-ion reduced mass and $E_\mathrm{col}$ is the collision energy~\cite{JPRBook}. For a Doppler cooled ion and our experimental parameters, $P_\mathrm{ODT}=0.9$~W, $T=5.6(0.2)\,\mu$K and $n_\mathrm{a}=2.1(0.3)\times 10^{16}\,$m$^{-3}$, we find $\Gamma_{3}<0.01\,$s$^{-1}$ which corresponds to the formation of one molecular ion in $<200$ experimental runs. Therefore, three-body recombination of molecular ions does not play a significant role in the explored parameter space. 

\begin{figure}
	\centering
	\includegraphics[width=1\linewidth]{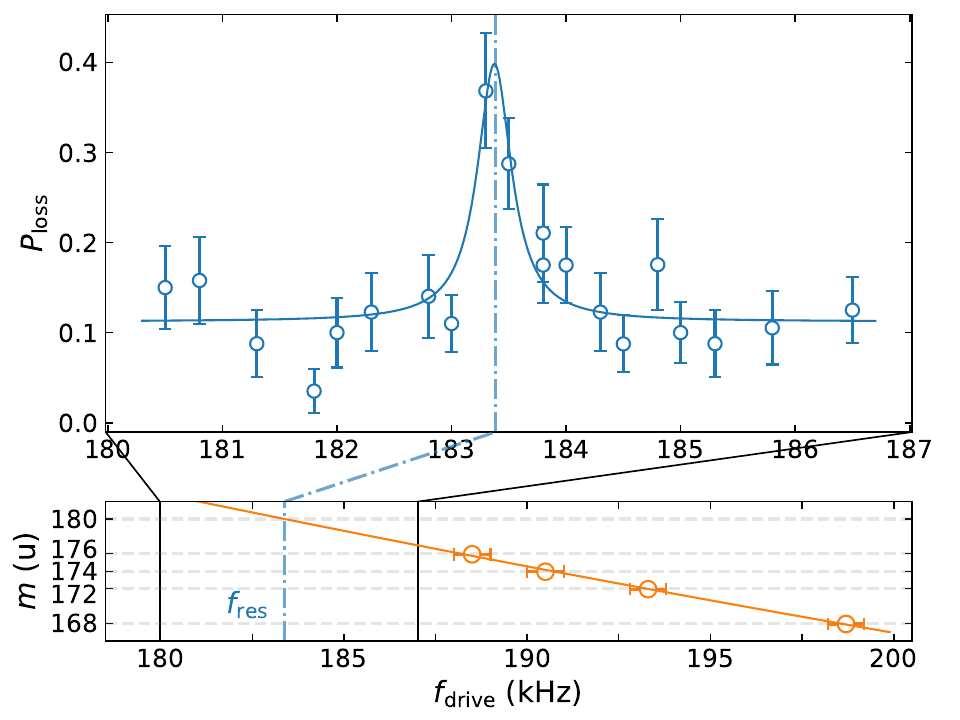}
	\caption{Mass spectrometry of LiYb$^+$ and Yb$^+$ by resonant driving the radial trap frequency with $f_\mathrm{drive}$. Top panel: ion loss (disks) with resonant driving applied during the interaction of $^{174}$Yb${^+}$ with an ultracold mixture of $^6$Li/Li$_2$, with
	each data point corresponding to at least 57 repetitions.
	The solid blue line is a Lorentzian fit to the data. The dash-dotted line indicates the expected trap frequency $f_\mathrm{res}\approx183.4\,$kHz for a single charged ion with mass number $180$. It is obtained from frequency calibration (bottom panel) using four Yb$^+$ isotopes~\cite{Suppl:2021}. The error bars reflect the projection noise.
}
 	\label{fig:trapFreq}
\end{figure}

To further investigate the observed probability of dark events, we perform mass spectrometry to detect the presence of LiYb$^+$, as shown in Fig.~\ref{fig:trapFreq}. 
An rf-electric field ($f_\mathrm{drive}=180-200\,$kHz) is applied to a cylindrical electrode (blue rod in Fig.~\ref{fig:setup}) from which energy can be transferred to the ion motion, when $f_\mathrm{drive}$ is in resonance with the ion's radial trap frequency. When enough energy is transferred to the ion, this process leads to ion loss as the particle is heated from the trap. Since the trap frequency of the ion depends on its mass, we can use this scheme for mass spectrometry. We calibrate our system and extract the expected trap frequency $f_\mathrm{res}$ for the molecular ion $m=180\,$u by measuring the trap frequencies of various isotopes of Yb$^+$~\cite{Suppl:2021}. 

Around $f_\mathrm{res}$ we find an enhancement of  Yb$^+$ ion loss (see Fig.~\ref{fig:trapFreq} (upper panel)), confirming that  LiYb$^+$ molecular ions are formed during the interaction of the Yb$^+$ with the atom-dimer bath. 
These measurements are done by applying a $2\,$Vpp driving during the entire sequence  and measuring ion loss ($t_\mathrm{c}>1500\,$ms)  as a function of $f_\mathrm{drive}$. For off-resonant frequencies, photo-dissociation of LiYb$^+$ in the ODT beams results in dark Yb$^+$ ions, that mainly return within $1.5\,$s of Doppler cooling and thus the background Yb$^+$ ion loss is low. Around $f_\mathrm{res}$ however, resonant heating moves  LiYb$^+$ quickly out of the ODT beams and reduces the photo-dissociation probability. Thus, driving LiYb$^+$ out of the ion trap results in increased loss events. For these measurements, we prepare the atom-dimer bath at
$T\approx 2\,\mu$K and $n_\mathrm{a}\approx 1\times 10^{16}\,$m$^{-3}$.

\begin{figure}
	\centering
	\includegraphics[width=\linewidth]{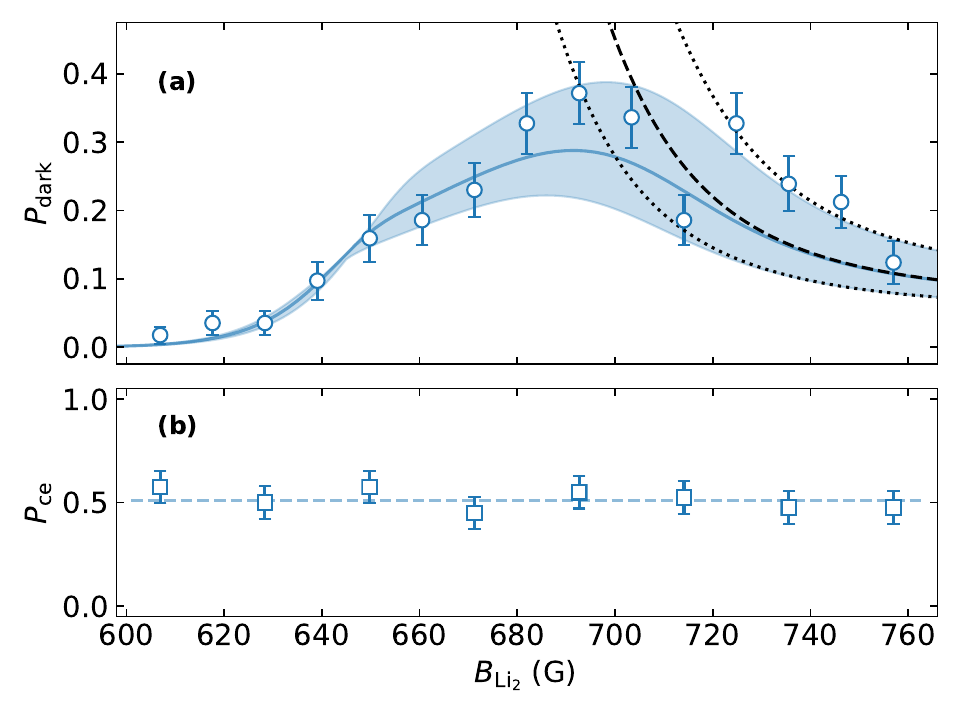}
 	\caption{a) Dark ion and b) charge exchange (ce) probabilities as a function of $B_\mathrm{Li_2}$.  Markers represent measurements, the black dashed line shows the simple thermal equilibrium model and the blue solid line shows the numerical solution to rate equations. Note, that no fitting parameters are used. The blue shaded region and the black dotted lines account for a $20\%$ error in the atom temperature, typical for time-of-flight measurements. The blue dashed line in b) shows the mean of $P_\mathrm{ce}$. The error bars show the total
statistical uncertainties.}
 	\label{fig:bScan}
\end{figure}

Finally, we study dimer-ion collisions in more detail by tweaking the dimer density in our system. The atomic three-body recombination rate for the process $\mathrm{Li} +\mathrm{Li}'+\mathrm{Li}\rightarrow \mathrm{Li_2}+\mathrm{Li}$ is a function of the atom scattering length, which can be tuned by the magnetic field $B_\mathrm{Li_2}$ in the vicinity of the $832\,$G Feshbach resonance. We use $P_\mathrm{ODT}\approx1.5\,$~W which results in $T=5.6(2)\,\mu$K and $n_\mathrm{a}=2.1(3) \times 10^{16}\,$m$^{-3}$ roughly constant over the explored magnetic field range. Note that the interactions with the ion always occur at $B=0\,$G to ensure the same binding energy during the collision with the ion. 

We find significant dark ion probabilities for $B_\mathrm{Li_2}$ approaching the Feshbach resonance. The measured $P_\mathrm{dark}$ as a function of $B_\mathrm{Li_2}$ is shown as blue disks in Fig.~\ref{fig:bScan} a), with each data point corresponding to 40 repetitions. Below $600\,$G, we find negligible dark ion probabilities, in agreement with Ref.~\cite{Feldker:2020}, whereas, when tuning $B_\mathrm{Li_2}$ above $600\,$G, we observe a significant increase of $P_\mathrm{dark}$, peaking around $693\,$G. For comparison, we measure the charge-exchange probability to obtain a relative local atom density with the results shown in panel (b) of Fig.~\ref{fig:bScan}. $P_\mathrm{ce}$ is approximately constant over the explored $B_\mathrm{Li_2}$, indicating that varying $B_\mathrm{Li_2}$ has no significant effect on the atomic density~\footnote{This is also confirmed by separate time-of-flight measurements of the atom density for varying $B_\mathrm{Li_2}$}.

The relationship between the probability of dark ions and the magnetic field can also be calculated via the dimer density $n_\mathrm{d}$. Two different approaches to obtain $n_\mathrm{d}$ result in the blue and the dashed black theory lines in Fig.~\ref{fig:bScan} a). The dimer density is related to the probability by $P_\mathrm{dark}(n_\mathrm{d})=1-e^{-\tau \lambda_\mathrm{d} n_\mathrm{d}}$, where $\tau=500\,$ms is the interaction time and $\lambda_\mathrm{d}=4.9\times10^{-15}\,$s$^{-1}$m$^{3}$ is the dimer-ion Langevin collision rate per unit density. Here, we rely on the fact that every dimer-ion collision results in a molecular ion when the collision energy is much smaller than the dimer binding energy. For the presented system, this was recently demonstrated with quasi-classical trajectory simulations~\cite{Hirzler2020a}. Moreover, the density of molecular states is much larger in the long range molecular ion potential than in the short range van der Waals potential of the dimer, i.e. it is more likely that a molecular ion is created. 

The dimer density $n_\mathrm{d}$ as a function of the magnetic field can be calculated by looking into the three-body recombination and dissociation process that determine the dimer formation~\cite{Chin2004, Kokkelmans2004, Suppl:2021}. Close to the Feshbach resonance, three-body recombination of Li$_2$ can be described with the rate coefficient $\propto T E_\mathrm{b}^{-3}$. This is in competition with three-body dissociation $\mathrm{Li_2}+\mathrm{Li}\rightarrow \mathrm{Li} +\mathrm{Li}'+\mathrm{Li}$ which has the rate coefficient $\propto T^{5/2}E_\mathrm{b}^{-3} \exp^{-E_\mathrm{b}/k_\mathrm{B} T}$, where $k_\mathrm{B}$ is the Boltzmann constant and $E_\mathrm{b}$ is the dimer binding energy~\cite{Chin2004}. We obtain the binding energy for different magnetic fields $B_\mathrm{Li_2}$ using precise measurements from Ref.~\cite{Zuern:2013}.

We find reasonable agreement with our ion-data for magnetic fields $\gtrsim700\,$G, when assuming an atom-dimer thermal equilibrium to obtain the dimer density. This is shown as the black dashed line in Fig.~\ref{fig:bScan}. Here, the dimer density follows an analytical expression as deduced in Refs.~\cite{Chin2004, Kokkelmans2004}, which we solve for the atom density and the atom temperatures of our system. However for fields below $700\,$G this simple model deviates from our measurements.

We find excellent agreement with the measured data over the entire magnetic field spectrum when numerically solving the rate equations as they evolve during the evaporation ramp~\cite{Suppl:2021}. This results in the blue solid lines in Fig.~\ref{fig:bScan}. It should be stressed that there are  no fitted parameters in our model.  The good agreement therefore indicates that every dimer-ion collision results in a molecular ion, which subsequently dissociates to a hot Yb$^+$ ion, observable as a dark event in the experiment.

The results show that we can use the single ion as a sensor for Li$_2$ dimers in our system. In particular, we probe the local dimer density as $n_\mathrm{d} = -\ln{(1-P_\mathrm{dark})}/(\tau \lambda_\mathrm{d})$. As an example, we consider $P_\mathrm{dark}=0.2$, which can easily be distinguished from the background (see Fig.~\ref{fig:bScan}). We find $n_\mathrm{d} \approx1\times 10^{14}\,$m$^{-3}$ corresponding to a relative density of $n_\mathrm{d}/n_\mathrm{a}\approx0.004$. Remarkably, this amounts to only about 50 Li$_2$ dimers in our atomic cloud and shows the potential of using trapped ions to detect trace amounts of molecular gases. 


\paragraph{Conclusion \& Outlook.}
\ \rere{We observed interactions of a single ion with ultracold Feshbach dimers and } identified Li$_2$+Yb$^+$ collisions as the origin for the \rere{created}
 LiYb$^+$ in our system. We found a strong correlation between molecular ion formation and the dimer density, and 
  observed molecular ions in our system via mass spectrometry. \rere{This is a new approach for creating molecular ions, which is independent of the mass ratio.}
\rere{It relies on the binding energy of the dimers being  larger than the ion-dimer collision energy and the absence of strong inelastic atom-ion loss reaction channels. Atom-ion collisions and three-body recombination could be  eliminated by purification of
the cloud using resonant laser pulses~\cite{Ferlaino:2009}.} Our results suggest the applicability of a single ion as a probe for trace molecule gases, with densities as low as $10^{14}\,$m$^{-3}$.  
\rere{This new technique might be used to detect Feshbach molecules of various atom combinations, whose quantities are  not detectable via commonly-used methods~\cite{Koehler2006, Ferlaino:2009, Julienne:2010}.}

Once the ion-dimer collision energy exceeds the Li$_2$ binding energy, dimer-dissociation should become prominent and reduce the molecular ion formation rate~\cite{Hirzler2020a}. \rere{This can be simply controlled with the magnetic field during the collision.} 
\hh{It will be interesting to study how the collision energy, in particular the micromotion, affects the crossover between the two regimes.} Feshbach dimers will allow studies of ultracold chemistry between ions and molecules, in which quantum effects such as ion-neutral Feshbach resonances~\cite{Tomza:2019, Weckesser:2021} will influence the reaction channels.

\nocite{Grimm2008ufg, Zwerger2012tbb, Petrov2003tbp, Ketterle1996eco, Moerdijk1995riu, Gribakin1993cot, Julienne2014, Bartenstein:2005}

\section{Acknowledgments}
We thank T. Feldker for suggesting this work, B. Pasquiou and S. Bennetts for support in exchanging the Li oven and B. Pasquiou, M. Borkowski, N.J. van Druten and R.J.C. Spreeuw for comments on the manuscript. This work was supported by the Dutch Research Council Start-up grant 740.018.008 (R.G.), Vrije Programma 680.92.18.05 (E.T., R.G., J.P.) and Quantum Software Consortium programme 024.003.037 (A.S.N.). R.S.L. acknowledges funding from the European Unions Horizon 2020 research and innovation programme under the Marie Sklodowska-Curie grant No 895473.

\section{Supplemental Material} 

\section{Ion trapping frequency calibration}

We measure resonant heating of various Yb$^+$ isotopes to calibrate the trapping frequency of our Paul trap. The ion's radial trap frequency in the $x$-direction can be expressed by \cite{Leibfried:2003}
\begin{align}
f_m &\equiv\omega_x/2\pi \approx \sqrt{\left( \frac{- k_\mathrm{x} Q U_\mathrm{dc} }{m y_0^2}  + \frac{{k^\prime_\mathrm{x}}^2 Q^2 U_\mathrm{rf}^2}{2 m^2 r_0^4\Omega^2} 
    \right)}\\
    &\equiv \sqrt{\frac{\kappa_\mathrm{x}}{m}+\frac{\kappa^\prime_\mathrm{x}}{m^2}}, 
    \label{eq:trapFreq}
\end{align}
with $Q$ the electric charge of an ion with mass $m$, trap parameters $r_0$ and $y_0$, $U_\mathrm{rf}$ and $U_\mathrm{dc}$ the radio-frequency (rf) and direct-current (dc) voltages respectively, and geometrical factors $k_\mathrm{x}$ and $k^\prime_\mathrm{x}$. Here,  the trap parameters are combined into new factors $\kappa_\mathrm{x}$ and $\kappa^\prime_\mathrm{x}$.
Without atoms, we measure $f_m$ for the isotopes $^{168}$Yb$^+$, $^{172}$Yb$^+$, $^{174}$Yb$^+$ and $^{176}$Yb$^+$, by resonantly driving the trap frequency with low driving voltage of $0.05\,$Vpp. Due to the Doppler shift, successful heating is visible by reduced ion fluorescence. The results are presented in Fig. 3 (lower panel) of the main text. We fit Eq.~\ref{eq:trapFreq} (orange line) to extract $\kappa_x$ and $\kappa'_x$ and use these to calculate a theoretical  trap frequency for the  $^6\mathrm{Li}^{174}\mathrm{Yb}^+$ molecular ion of $f_\mathrm{res}\approx183.4\,$kHz (blue dash-dotted line). Note, that we observe day-to-day frequency drifts on the order of $0.5\,$kHz which we compensate by daily referencing to $f_{174}$.

\section{Observation of dark ions}
 
We observe dark ions after the interaction with the neutral cloud. We explain their origin in photo-dissociation of molecular ions with the $1064\,$nm ODT light, similar to observations with BaRb$^+$ reported in Ref.~\cite{Mohammadi2021}. 
\begin{figure}
	\centering
	\includegraphics[width=1\linewidth]{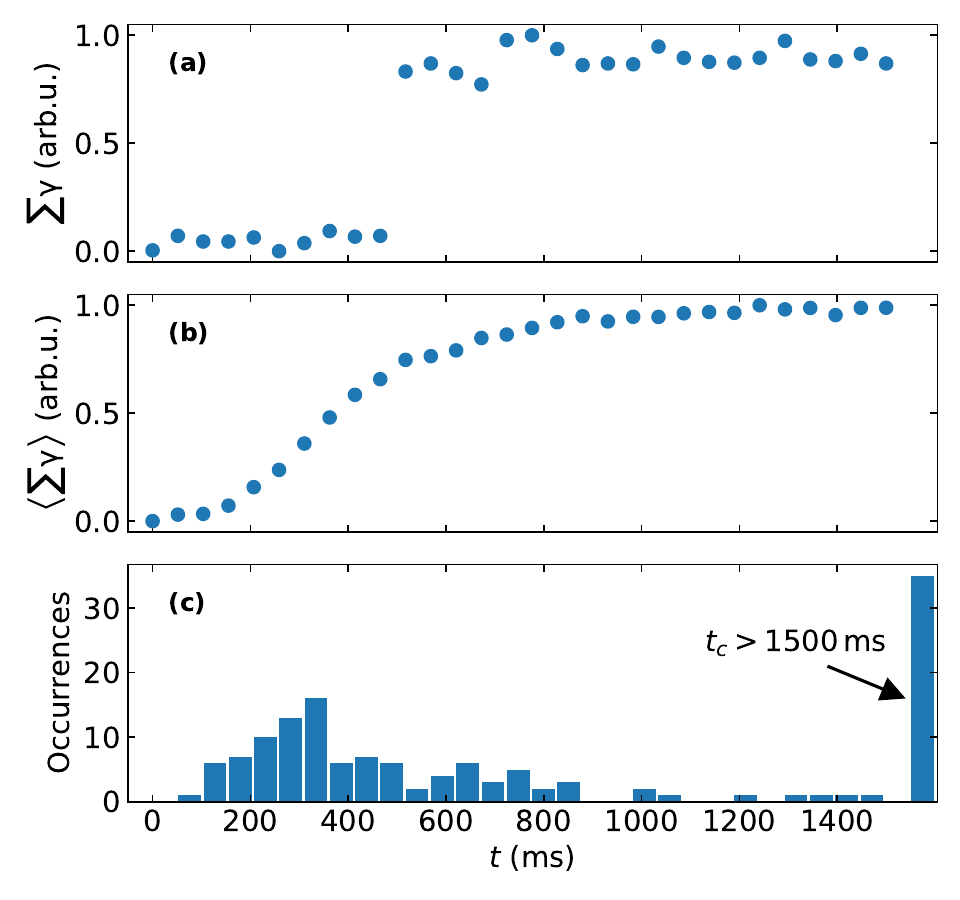}
 	\caption{Photon count ($\Sigma \gamma$) during Doppler cooling after an ion-neutral collision averaged over $2.5\,$ms detection time. (a) Single-shot photon count for a dark ion. 
 	(b) Average photon count and (c) histogram for all dark ions observed in scanning the ODT power in the main text.
 	}
 	\label{fig:recool}
\end{figure}
After the interaction with the cloud, we record fluorescence with a photomultiplier tube (pmt) every $50\,$ms over $1500\,$ms, whereby the detection time for each data point is $2.5\,$ms. For each fluorescence measurement, we project the photon count to represent a bright (fluorescent) ion or a dark (non-fluorescent) ion. Then, we combine the photon counts from all measurements into a histogram and we define a threshold between the partially separated peaks for bright and dark ions. Varying the threshold by $20\,\%$, we do not find a significant deviation on our results. 

For the $P_\mathrm{ODT}$ scan data (Fig.2) of the main text, Fig~\ref{fig:recool}a) shows a typical single shot photon count for a dark ion turning bright again after about $500\,$ms of Doppler cooling, together with in b) the fluorescence averaged over the entire data set.  The same data is shown as a histogram in c), where the contribution from lost ions ($t_\mathrm{c}>1500\,$ms) can be seen. 

\section{Creation of dimers and density model}
The dimers are created by three-body recombination during evaporation close to the 832\,G Feshbach resonance (FR). This particular resonance has been widely studied (e.g.~\cite{Grimm2008ufg, Zwerger2012tbb}) and we can use the insights from~\cite{Chin2004, Kokkelmans2004} to calculate the dimer density in our system.

\subsection{Rate Equations}
For a positive scattering lengths ($B<832\,$G) and a sample initially consisting of atoms, three-body recombination is the dominant process to form dimers,  
\begin{equation}
\text{Li}+\text{Li}'+\text{Li}\rightarrow\text{Li}_2+\text{Li}.
\end{equation}
The dissociation process is the other way around,
\begin{equation}
\hh{\text{Li}_2+\text{Li}\rightarrow\text{Li}+\text{Li}'+\text{Li},}
\end{equation}
 and the two spinstates of lithium are indicated by $\text{Li}$ and $\text{Li}'$.  The formation rate coefficient is given by $R_{3r}=167a^6k_\mathrm{B}T/\hbar$ when $E_\mathrm{b}/k_\mathrm{B}T \gg 1$~\cite{Petrov2003tbp}, with $E_\mathrm{b}$ the binding energy of the dimer. Here, $a$ is the atom scattering length, $k_\mathrm{B}$ the Boltzmann constant, $T$ the atom temperature and $\hbar$ Planck's constant $h$ divided by $2\pi$. In the halo-regime where $E_\mathrm{b}/h< E_\mathrm{vdW}/h=614\,$MHz, \hh{with van der Waals energy $E_\mathrm{vdW}$}, 
this results in $R_{3r}\approx 167 \frac{\hbar^5}{m_\mathrm{a}^3}\frac{k_\mathrm{B} T}{E_\mathrm{b}^3}$ and the dissociation rate $C_{d}=3.75\frac{\hbar^2}{m_\mathrm{a}^{3/2}}\frac{(k_\mathrm{B}T)^{5/2}}{E_\mathrm{b}^3} e^{-E_\mathrm{b}/(k_\mathrm{B}T)}$ for the reverse process~\cite{Chin2004}. This leads to the following rate equations 

\begin {equation}
\frac{d n_\mathrm{a}}{dt}= - 4R_{3r}n_\mathrm{a}^3 + 4C_{d}n_\mathrm{a}n_\mathrm{d}
\label{Eq:na}
\end {equation}
\begin {equation}
\frac{d n_\mathrm{d}}{dt}= 2 R_{3r}n_\mathrm{a}^3 - 2 C_{d}n_\mathrm{d}n_\mathrm{a}.
\end {equation}
 where $n_\mathrm{a}$ is the atom density of a single spin state and $n_\mathrm{d}$ the dimer density. Here, we assume that all atoms and dimers remain trapped even after colliding and thus that $E_\mathrm{b}\ll U_\mathrm{trap}$, the trap depth which is about $20(14)\,\mu$K.

 \subsubsection{Thermal Equilibrium}
 In thermal equilibrium, $\frac{d n_\mathrm{a}}{dt}=\frac{d n_\mathrm{d}}{dt}=0$, and assuming the atoms and molecules thermalize ($T=T_\mathrm{a}=T_\mathrm{d}$), Eq.~\ref{Eq:na} simplifies to an analytic expression
\begin{equation}
\label{Eqnd}
n_\mathrm{d}=
\frac{R_{3r}}{C_{d}} n_\mathrm{a}^2=2^{3/2}\left(\frac{2\pi \hbar^2}{m_\mathrm{a} k_\mathrm{B} T_\mathrm{a}}\right)^{3/2}\,  n_\mathrm{a}^2 \, e^{E_\mathrm{b}/(k_\mathrm{B}T)}.
\end{equation}
In terms of phase space density, the equation can be rewritten as $\phi_\mathrm{d}=\phi_\mathrm{a}^2\,e^{E_\mathrm{b}/(k_\mathrm{B}T)}$ , with  $\phi_\mathrm{d} (\phi_\mathrm{a})$ the dimer(atom) phase space density, respectively, as found in~\cite{Chin2004}.

The results for this simple analytic model are shown in Fig. 4 of the main text as dashed black lines. Agreement is found for magnetic fields above $700\,$G. For lower magnetic fields the binding energy of the dimers is greater than $10\,\mu$K and the assumptions that all particles during three-body recombination and dissociation remain trapped is no longer valid. 

\subsubsection{Including Evaporation}
For an accurate picture of the dimer formation, the evaporation ramp needs to be taken into account as our dimers are created during the last stage of evaporation. The evaporation ramp changes the rate equations because the rate coefficients now become time-dependent, as the temperature of the atom cloud changes with time. Furthermore, there is an additional loss channel for the atoms as they evaporate out of the trap. As atoms evaporate and the temperature of the cloud changes, the rate coefficients change according to, i.e. $ R_{3r}\propto T_\mathrm{a}(t)$ and $C_{d}\propto\left[T_\mathrm{a}(t)\right]^{5/2}\,e^{-E_\mathrm{b}/(k_\mathrm{B} T_\mathrm{a}(t))}$, respectively.

During the evaporation ramp, the change in trapdepth $U$ and atomnumber $N$, leads to a change in atom density which can be characterized by
\begin{equation}
\gamma(t)=\left[(1-\frac{3}{2}\alpha)\frac{\dot{N}}{N(t)}+3\frac{\dot{U}}{U(t)}\right] n_\mathrm{a},
\end{equation}
Here, $\alpha$ is the scaling parameter, whereby $T\propto N^\alpha$, and represents the temperature decrease per lost particle~\cite{Ketterle1996eco}. We assume an exponential evaporation ramp of duration $\tau_\mathrm{e}$. Thus $N(t)= N_\mathrm{i}\, e^{\left(\frac{t}{\tau_\mathrm{e}}\, \text{Ln}\left[\frac{N_\mathrm{f}}{N_\mathrm{i}}\right]\right)} $ and $U(t)= U_\mathrm{i}\, e^{\left(\frac{t}{\tau_\mathrm{e}}\,\text{Ln}\left[\frac{U_\mathrm{f}}{U_\mathrm{i}}\right]\right)}$, with $N_\mathrm{i}(N_\mathrm{f})$ the initial (final) atom number and $U_\mathrm{i}(U_\mathrm{f})$ the initial (final) trap depth at the beginning (end) of the evaporation ramp.

The full rate equations now become
\begin {equation}
\frac{d n_\mathrm{a}}{dt}=-4 R_{3r} n_\mathrm{a}^3+ 4 C_{d} n_\mathrm{d}n_\mathrm{a} -\gamma(t)
\end {equation}
\begin {equation}
\frac{d n_\mathrm{d}}{dt}= 2 R_{3r}n_\mathrm{a}^3 - 2 C_{d}n_\mathrm{d}n_\mathrm{a}.
\end {equation}
The results of solving these rate equations for our experimental parameters are shown as the blue line in Fig. 4 of the main text. Note that no free parameters are used and the experimental input parameters are obtained from time-of-flight (tof) data. See table~\ref{tab:atom} for the simulation input values and corresponding statistical uncertainities, as obtained from tof absorption images. The values for $N_\mathrm{i}, T_\mathrm{i}, n_\mathrm{a} (t=0), U_\mathrm{i}, U_\mathrm{f}$ and $\alpha$ are based on our evaporation ramp data, taken separately and rescaled to match the tof data of the ion-measurement day. For calculating the trap depth from the tof data, we use $U=\frac{\bar{w}_0^2}{4}\left(\sigma_\mathrm{x}^0\sigma_\mathrm{y}^0\sigma_\mathrm{z}^0\right)^{-
2/3} k_\mathrm{B}T$, with average width $\bar{w}_0=\left(w_\mathrm{x}\,w_\mathrm{y}\,w_\mathrm{z}\right)^{1/3}\approx \left(40\times40\times4000\right)\,\mu$m $ \approx{186}\,\mu$m and $\sigma_\mathrm{z}^0\approx 10\,\sigma_\mathrm{x}^0$. Here, $\sigma_{i}^0$ is the in-situ width of the cloud in the $i$-th direction with $i=\left(\mathrm{x},\mathrm{y},\mathrm{z}\right)$. From the atom tof data we obtain both $\sigma_\mathrm{x}^0$ and $\sigma_\mathrm{y}^0$ and the trap ratio (1:10) gives the relation between $\sigma_\mathrm{z}^0$ and $\sigma_\mathrm{x}^0$. 
The final atomnumber $N_\mathrm{f}$ comes from the tof taken in between the ion-measurements on the same day. 
We check that the numerical simulation of the evaporation results in the final density and temperature as measured in these time-of-flights. The mean values for these datasets are $n_\mathrm{a}=2.1(3) \times 10^{16}\,$m$^{-3}$ and $T=5.6(2)\,\mu$K.

\begin{table}[t]
    \centering
    \begin{tabular}{|c|c|}\hline
	   Parameter & Value \\\hline
            $N_\mathrm{i}$ & $17(3)\times 10^4$  \\ $N_\mathrm{f}$ & $33(2)\times 10^3$ \\
           $T_\mathrm{i}$ ($\mu$K) &37(8)  \\ $\alpha$ & 1.1(2) \\
          $n_\mathrm{a}^{(t=0)}$ (\,m$^{-3}$) & $5(2)\times 10^{16}$ \\\hline
          $U_\mathrm{i}$ ($\mu$K) & 73(54)  \\ $U_\mathrm{f}$  ($\mu$K)  & 20(14) \\\hline
    \end{tabular}
    \caption{Values of atom parameters used in the numerical solutions to the rate equations (solid lines, Fig.~4. main text) and based on time-of-flight (tof) absorption imaging. Errors come from the fit uncertainity of the parameters extracted from the tof fits. }
    \label{tab:atom}
\end{table}

\subsection{Binding energy of the dimers}
In the universal regime of the Feshbach resonance ($|a|\gg R_\mathrm{vdW}$), the binding energy is given by $
E_\mathrm{b}=\hbar^2/(m a^2) $, with $m$ the mass of the atom and $R_\mathrm{vdW}$ the van der Waals range. The scattering length $a$ follows the general Feshbach relation~\cite{Moerdijk1995riu, Chin:2010} of $a=a_\text{bg}-\frac{a_\text{bg}\Delta}{B-B_0}$. This shows that, by changing the magnetic field, we can tune the scattering length and change the binding energy of the dimers.  

When calculating the binding energy for magnetic fields further away from the center, such that $a\gg R_\mathrm{vdW}$ does not hold, an additional correction to the universal expression of $E_\mathrm{b}$ needs to be taken into account, which comes from the non-zero range of the van der Waals potential~\cite{Gribakin1993cot}. Then
\begin{equation}
E_\mathrm{b}=\frac{\hbar^2}{m(a-\bar{a})^2}.
\label{Ebbar}
\end{equation}
For the case of lithium $\bar{a}=0.956 R_\text{vdW}= 0.956\times 31.26 a_0$ according to~\cite{Grimm2008ufg}. The dimers are created in the $^3\Sigma_u^+ (\nu=-1)$ state, where $\nu$ is the vibrational quantum number counted down from the continuum~\cite{Chin:2010}. This is the last molecular bound state of Li$_2$ which mixes with the entrance channel $(F=1/2, m_\mathrm{F}=\pm 1/2)$ facilitating formation of Li$_2$ dimers.

Note that when we overlap the atoms with the ion, we ramp to zero-field to perform all experiments with the same dimer binding energy to avoid influences on the molecular ion formation rate~\cite{Hirzler2020a} as well as quantum effects~\cite{Tomza:2019, Weckesser:2021}. For magnetic fields below $550\,$G the dimer bound state switches from $^3\Sigma_u^+ (\nu=-1)$  to $^1\Sigma_g^+ (\nu=-1)$ and with $S=0$ its magnetic moment vanishes. Consequently, the binding energy of the dimers becomes field independent and for zero-field they have a binding energy of about $1.38\,$GHz~\cite{Chin:2010, Julienne2014}. 

\subsection{Feshbach Resonance Parameters}
The exact value of $E_\mathrm{b}(B)$ or $a(B)$ matters when we want to calculate the number of dimers we create. This value is closely related to the determination of the FR parameters of the FR we use to create the dimers. The most accurate and recent determination of the FR parameters stems from ~\cite{Zuern:2013}, where they measured the binding energy as a function of magnetic field directly and provide $a(B)$ for 0-2000\,G. When doing a local fit ($|B-B_0|\ll \Delta$) to $a(B)$ using $a=a_\text{bg}-\frac{a_\text{bg}\Delta}{B-B_0}$, they found the fit parameters to be $a_\text{bg}=-1582(1)\,a_0$, $B_0=832.18(8)\,G$, $\Delta=-262.3(3)\,G$. However this only applies for a narrow range of magnetic fields close to $B_0$. Farther away, a leading order correction term needs to be added because of another close-by FR at 527\,G, as was shown in~\cite{Bartenstein:2005}. For our theory lines we therefore choose to directly take $a(B)$ as given by~\cite{Zuern:2013} and use Eq.~\ref{Ebbar} to calculate the binding energy.

\subsection{Tuning the dimer density}
\begin{figure}
	\centering
	\includegraphics[width=1\linewidth]{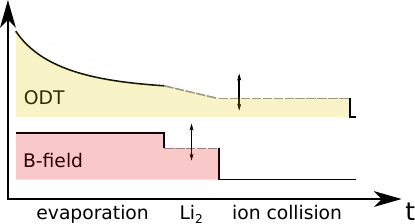}
 	\caption{Sketch of evaporation sequence to tune the dimer density in Li-Li$_2$ mixture before overlap with a single ion. We either vary the magnetic field or the trap depth of the final evaporation stage.}
 	\label{fig:evapSeq}
\end{figure}

To explore different Li$_2$ densities in our system, we either tune $P_\mathrm{ODT}$ or $B_\mathrm{Li_2}$ at the final stage of the evaporative cooling in the experimental sequence, as it is depicted in Fig.~\ref{fig:evapSeq}. We tune the atoms' phase space density, by changing the ODT depth at the end of evaporation with the help of an acousto-optical modulator. This changes both the density of the atoms as well as their temperature.  The $P_\mathrm{ODT}$ is the power of the optical dipole beam when entering the setup. This beams creates the crossed dipole trap by entering through the endcaps of the ion trap at an angle of 5$^{\circ}$ with respect to the Paul trap axis as depicted in Fig.1 of the main text. Using a lin $\perp$ lin polarization configuration, we prevent the occurrence of an optical lattice potential. 

When changing the magnetic field to {$B_\mathrm{Li_2}=600-760\,$G}, we found that the atom density stays constant as the amount of dimers created stays below the experimental error (about 20) for determining the atom density. This is further confirmed by the constant charge exchange rate measured by the ion and from independent time-of-flight measurements. When changing $B_\mathrm{Li_2}$, the scattering length changes and thus the dimer binding energy. This leads to a variable dimer density as the recombination rate and dissociation rate are depending on $E_\mathrm{b}$.

\section{Atom Data for $P_\mathrm{ODT}$ scan}
We take regular time-of-flight measurements of the atomic cloud using absorption imaging after we transport the ODT to the center of the Paul trap. We typically average 4-5 images before fitting a Gaussian distribution to extract the atom parameters, i.e. atom number and width of the cloud. The coldest atom clouds have a temperature as low as $0.9(5)\,\mu$K with about $5\times10^3$ atoms. However the presence of the ion trap limits the numerical apperture of our imaging system and the coldest clouds are only a few pixels wide and  when expanding become quickly to dilute too detect. 

From the time-of-flight curves, assuming free expansion, the atomnumber $N _\mathrm{a}$, temperature $T_x$ and $T_y$ as well as the in-situ width of the cloud, $\sigma^0_x$ and $\sigma^0_y$,  in both x- and y- direction can be obtained. This follows from fitting the expansion of the width of the cloud as a function of expansion time using
\begin{equation}
\sigma_i (t)=\sqrt{(\sigma^0_i)^2 + \frac{k_B T}{m_a}\,t^2}
\end{equation}
whereby the in-situ width of the cloud is related to the trapfrequency by $\sigma^0_i=\left( \frac{k T}{m \omega_i^2}\right)^{1/2}$, with trap frequencies $\omega_i$ and $i=(\mathrm{x},\mathrm{y},\mathrm{z})$. 

The peak density $n_\mathrm{a}$ for a thermal cloud in an harmonic trap is given by
\begin{equation}
n_\mathrm{a}=\left(\frac{\bar{\omega}^2\,m_\mathrm{a}}{2\pi\, k_\mathrm{B}\, T_a}\right)^{3/2}N= \frac{N}{(2\pi)^{3/2} \sigma^0_\mathrm{x}\, \sigma^0_\mathrm{y}\, \sigma^0_\mathrm{z}}
\end{equation}
Here, $\bar{\omega}^2$ is the geometrical average of the trap frequency and it can be calculated as $\bar{\omega}_{i}=(\omega_\mathrm{x}\omega_\mathrm{y}\omega_\mathrm{z})^{1/3}$. Furthermore $T=\left(T_\mathrm{x} +T_\mathrm{y}\right)/2$ and $\sigma_\mathrm{z}^0=10\,\sigma_\mathrm{x}^0$ for our trap. The phase space density is given by 
\begin{equation}
    \phi_\mathrm{a}=\hh{n_\mathrm{a}}\,\lambda_\mathrm{dB}^3= \hh{n_\mathrm{a}}\left(\frac{2\pi \hbar^2}{m_\mathrm{a} k_\mathrm{B} T_\mathrm{a}}\right)^{3/2}.
\end{equation}

\begin{figure}
	\centering
	\includegraphics[width=1\linewidth]{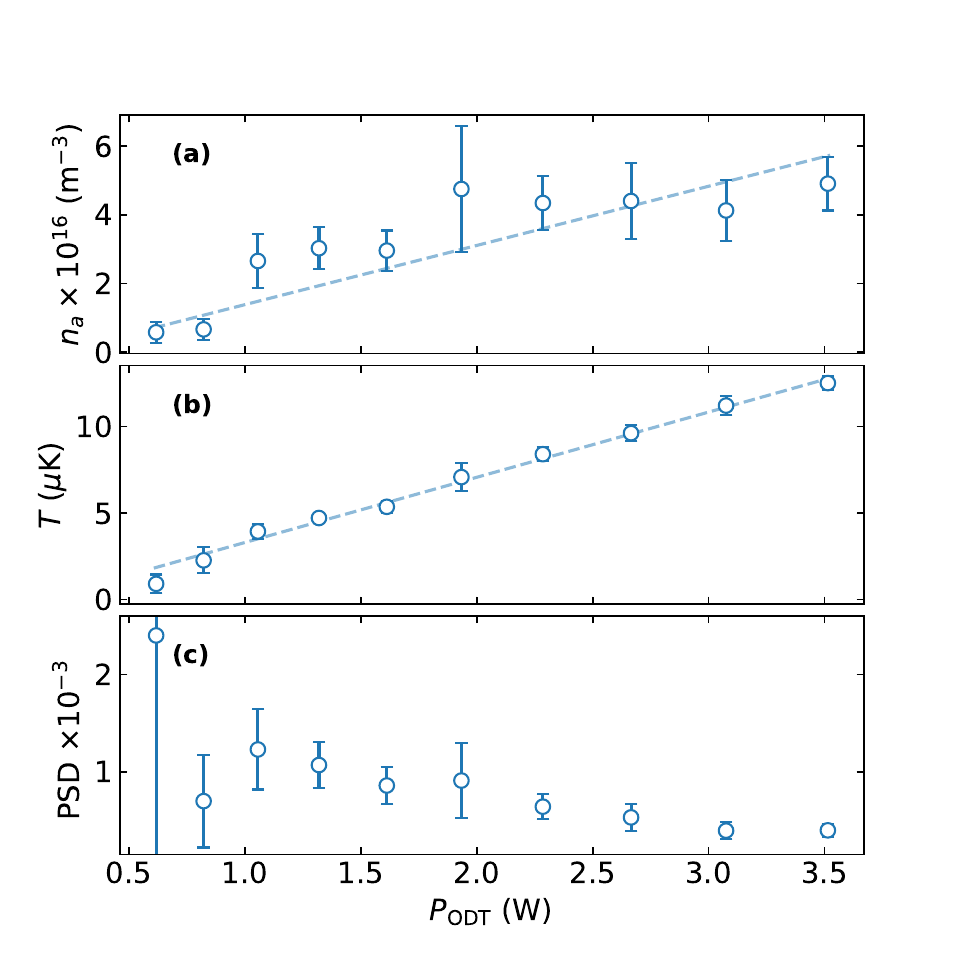}
 	\caption{Atom parameters for optical dipole trap power scan. The atom density a), temperature b) and phase space density (PSD) c) are obtained from time-of-flight absorption imaging measurements. The dashed lines are linear fits and guides for the eye. The atom PSD in c) is related to the increase in dimer phase space density. Here, the error bars reflect the total statistical uncertainties. Note the high error bars for the lowest point ($P_\mathrm{ODT}\approx 0.7$\,W) in c), showing the limitations of our imaging method.}
 	\label{fig:tof}
\end{figure}

 For the $P_\mathrm{ODT}$ scan, the observed atom density $n_\mathrm{a}$, temperature $T$ and phase space density $\phi_\mathrm{a}$ are shown in Fig.~\ref{fig:tof}.

 \rere{We can model the dimer density using the thermal equilibrium model (Eq.~\ref{Eqnd}) based on the atom data
corresponding to Fig. 2b (main text), which only requires the final atom number and temperature as input. This model is shown in Fig. 4 (black dashed line) with a 20\% error range in atom temperature (black dotted lines). The model confirms the trend we see in the ion-measurements. Furthermore, the overestimation of the dark ion
probability is similar to the result in Fig. 4a (main text), where the ion-measurement at 693\,G is also overestimated by the black curve. Further benchmarking of the evaporation ramps used in this dataset would allow us to include the evaporation ramp and numerically model the dimer density. This requires additional measurements of the initial atom number, temperature and density at the start of the evaporation ramp, as well as the initial trapdepth. Furthermore, for each datapoint we varied the final trapdepth and this affects the scaling parameter $\alpha$ that describes the
evaporation and this would have to be individually benchmarked as well. Nevertheless the simple thermal equilibrium model can qualitatively explain the observed increase in the probability of the molecular ion formation for lower $P_\mathrm{odt}$ as observed in the main text (Fig. 2).
}

 
 \begin{figure}
	\centering
	\includegraphics[width=1\linewidth]{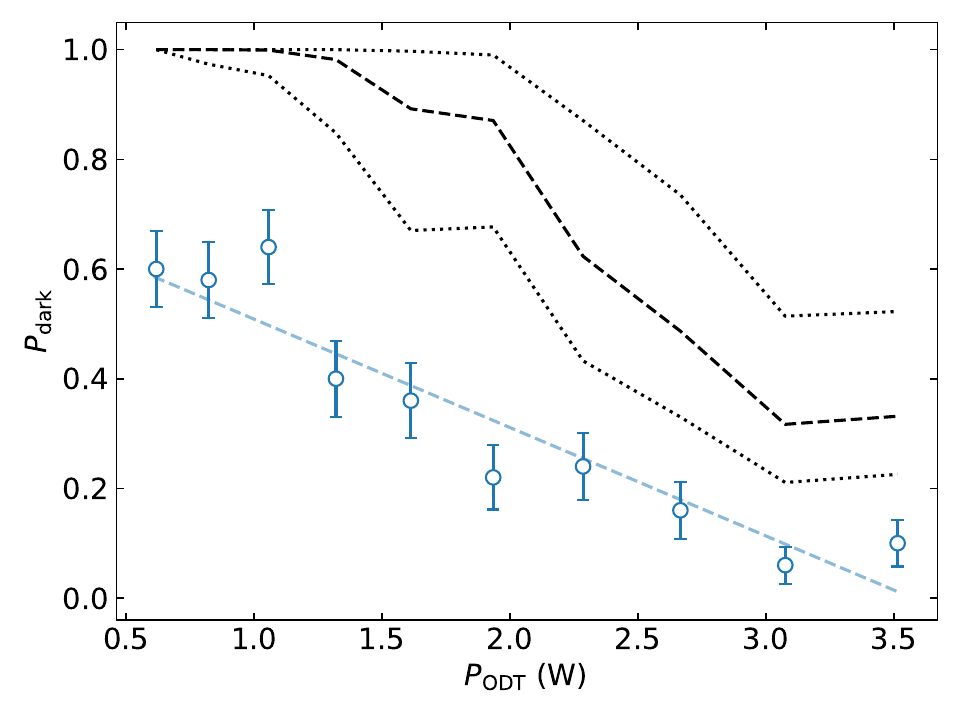}
 	\caption{\rere{Dark Yb$^+$ ion probability as a function of the ODT power. The experimental data (blue disks) is the same as in Fig~2 a) of the main text and the black lines are the solution of Eq.~\ref{Eqnd} for the given experimental parameters and assuming thermal equilibrium. The black dotted lines account for a 20\% error in the atom temperature, typical for time-of-flight measurements. }}
 	\label{fig:evap2}
\end{figure}

%


\begin{thebibliography}{53}%
\makeatletter
\providecommand \@ifxundefined [1]{%
 \@ifx{#1\undefined}
}%
\providecommand \@ifnum [1]{%
 \ifnum #1\expandafter \@firstoftwo
 \else \expandafter \@secondoftwo
 \fi
}%
\providecommand \@ifx [1]{%
 \ifx #1\expandafter \@firstoftwo
 \else \expandafter \@secondoftwo
 \fi
}%
\providecommand \natexlab [1]{#1}%
\providecommand \enquote  [1]{``#1''}%
\providecommand \bibnamefont  [1]{#1}%
\providecommand \bibfnamefont [1]{#1}%
\providecommand \citenamefont [1]{#1}%
\providecommand \href@noop [0]{\@secondoftwo}%
\providecommand \href [0]{\begingroup \@sanitize@url \@href}%
\providecommand \@href[1]{\@@startlink{#1}\@@href}%
\providecommand \@@href[1]{\endgroup#1\@@endlink}%
\providecommand \@sanitize@url [0]{\catcode `\\12\catcode `\$12\catcode
  `\&12\catcode `\#12\catcode `\^12\catcode `\_12\catcode `\%12\relax}%
\providecommand \@@startlink[1]{}%
\providecommand \@@endlink[0]{}%
\providecommand \url  [0]{\begingroup\@sanitize@url \@url }%
\providecommand \@url [1]{\endgroup\@href {#1}{\urlprefix }}%
\providecommand \urlprefix  [0]{URL }%
\providecommand \Eprint [0]{\href }%
\providecommand \doibase [0]{http://dx.doi.org/}%
\providecommand \selectlanguage [0]{\@gobble}%
\providecommand \bibinfo  [0]{\@secondoftwo}%
\providecommand \bibfield  [0]{\@secondoftwo}%
\providecommand \translation [1]{[#1]}%
\providecommand \BibitemOpen [0]{}%
\providecommand \bibitemStop [0]{}%
\providecommand \bibitemNoStop [0]{.\EOS\space}%
\providecommand \EOS [0]{\spacefactor3000\relax}%
\providecommand \BibitemShut  [1]{\csname bibitem#1\endcsname}%
\let\auto@bib@innerbib\@empty
\bibitem [{\citenamefont {Sugiyama}\ and\ \citenamefont
  {Yoda}(1995)}]{Sugiyama_1995}%
  \BibitemOpen
  \bibfield  {author} {\bibinfo {author} {\bibfnamefont {K.}~\bibnamefont
  {Sugiyama}}\ and\ \bibinfo {author} {\bibfnamefont {J.}~\bibnamefont
  {Yoda}},\ }\href {\doibase 10.1143/jjap.34.l584} {\bibfield  {journal}
  {\bibinfo  {journal} {Jpn. J. Appl. Phys.}\ }\textbf {\bibinfo {volume}
  {34}},\ \bibinfo {pages} {L584} (\bibinfo {year} {1995})}\BibitemShut
  {NoStop}%
\bibitem [{\citenamefont {Sugiyama}\ and\ \citenamefont
  {Yoda}(1997)}]{Sugiyama1997}%
  \BibitemOpen
  \bibfield  {author} {\bibinfo {author} {\bibfnamefont {K.}~\bibnamefont
  {Sugiyama}}\ and\ \bibinfo {author} {\bibfnamefont {J.}~\bibnamefont
  {Yoda}},\ }\href {\doibase 10.1103/physreva.55.r10} {\bibfield  {journal}
  {\bibinfo  {journal} {Phys. Rev. A}\ }\textbf {\bibinfo {volume} {55}},\
  \bibinfo {pages} {R10(R)} (\bibinfo {year} {1997})}\BibitemShut {NoStop}%
\bibitem [{\citenamefont {Hoang}\ \emph {et~al.}(2020)\citenamefont {Hoang},
  \citenamefont {Jau}, \citenamefont {Overstreet},\ and\ \citenamefont
  {Schwindt}}]{Hoang2020}%
  \BibitemOpen
  \bibfield  {author} {\bibinfo {author} {\bibfnamefont {T.~M.}\ \bibnamefont
  {Hoang}}, \bibinfo {author} {\bibfnamefont {Y.-Y.}\ \bibnamefont {Jau}},
  \bibinfo {author} {\bibfnamefont {R.}~\bibnamefont {Overstreet}}, \ and\
  \bibinfo {author} {\bibfnamefont {P.~D.~D.}\ \bibnamefont {Schwindt}},\
  }\href {\doibase 10.1103/PhysRevA.101.022705} {\bibfield  {journal} {\bibinfo
   {journal} {Phys. Rev. A}\ }\textbf {\bibinfo {volume} {101}},\ \bibinfo
  {pages} {022705} (\bibinfo {year} {2020})}\BibitemShut {NoStop}%
\bibitem [{\citenamefont {Rugango}\ \emph {et~al.}(2015)\citenamefont
  {Rugango}, \citenamefont {Goeders}, \citenamefont {Dixon}, \citenamefont
  {Gray}, \citenamefont {Khanyile}, \citenamefont {Shu}, \citenamefont
  {Clark},\ and\ \citenamefont {Brown}}]{Rugango2015}%
  \BibitemOpen
  \bibfield  {author} {\bibinfo {author} {\bibfnamefont {R.}~\bibnamefont
  {Rugango}}, \bibinfo {author} {\bibfnamefont {J.~E.}\ \bibnamefont
  {Goeders}}, \bibinfo {author} {\bibfnamefont {T.~H.}\ \bibnamefont {Dixon}},
  \bibinfo {author} {\bibfnamefont {J.~M.}\ \bibnamefont {Gray}}, \bibinfo
  {author} {\bibfnamefont {N.~B.}\ \bibnamefont {Khanyile}}, \bibinfo {author}
  {\bibfnamefont {G.}~\bibnamefont {Shu}}, \bibinfo {author} {\bibfnamefont
  {R.~J.}\ \bibnamefont {Clark}}, \ and\ \bibinfo {author} {\bibfnamefont
  {K.~R.}\ \bibnamefont {Brown}},\ }\href
  {https://iopscience.iop.org/article/10.1088/1367-2630/17/3/035009/meta}
  {\bibfield  {journal} {\bibinfo  {journal} {J. Phys.}\ }\textbf {\bibinfo
  {volume} {17}},\ \bibinfo {pages} {035009} (\bibinfo {year}
  {2015})}\BibitemShut {NoStop}%
\bibitem [{\citenamefont {Willitsch}\ \emph {et~al.}(2008)\citenamefont
  {Willitsch}, \citenamefont {Bell}, \citenamefont {Gingell}, \citenamefont
  {Procter},\ and\ \citenamefont {Softley}}]{Willitsch:2008}%
  \BibitemOpen
  \bibfield  {author} {\bibinfo {author} {\bibfnamefont {S.}~\bibnamefont
  {Willitsch}}, \bibinfo {author} {\bibfnamefont {M.~T.}\ \bibnamefont {Bell}},
  \bibinfo {author} {\bibfnamefont {A.~D.}\ \bibnamefont {Gingell}}, \bibinfo
  {author} {\bibfnamefont {S.~R.}\ \bibnamefont {Procter}}, \ and\ \bibinfo
  {author} {\bibfnamefont {T.~P.}\ \bibnamefont {Softley}},\ }\href
  {https://journals.aps.org/prl/abstract/10.1103/PhysRevLett.100.043203}
  {\bibfield  {journal} {\bibinfo  {journal} {Phys. Rev. Lett.}\ }\textbf
  {\bibinfo {volume} {100}},\ \bibinfo {pages} {043203} (\bibinfo {year}
  {2008})}\BibitemShut {NoStop}%
\bibitem [{\citenamefont {R\'ios}(2020)}]{JPRBook}%
  \BibitemOpen
  \bibfield  {author} {\bibinfo {author} {\bibfnamefont {J.~P.}\ \bibnamefont
  {R\'ios}},\ }\href@noop {} {\emph {\bibinfo {title} {An introduction to cold
  and ultracold chemistry}}}\ (\bibinfo  {publisher} {Springer},\ \bibinfo
  {address} {Cham, Switzerland},\ \bibinfo {year} {2020})\BibitemShut {NoStop}%
\bibitem [{\citenamefont {Heazlewood}\ and\ \citenamefont
  {Softley}(2021)}]{Heazlewood2021}%
  \BibitemOpen
  \bibfield  {author} {\bibinfo {author} {\bibfnamefont {B.~R.}\ \bibnamefont
  {Heazlewood}}\ and\ \bibinfo {author} {\bibfnamefont {T.~P.}\ \bibnamefont
  {Softley}},\ }\href {\doibase 10.1038/s41570-020-00239-0} {\bibfield
  {journal} {\bibinfo  {journal} {Nat. Rev. Chem.}\ }\textbf {\bibinfo {volume}
  {5}},\ \bibinfo {pages} {125} (\bibinfo {year} {2021})}\BibitemShut {NoStop}%
\bibitem [{\citenamefont {Köhler}\ \emph {et~al.}(2006)\citenamefont
  {Köhler}, \citenamefont {G{\'{o}}ral},\ and\ \citenamefont
  {Julienne}}]{Koehler2006}%
  \BibitemOpen
  \bibfield  {author} {\bibinfo {author} {\bibfnamefont {T.}~\bibnamefont
  {Köhler}}, \bibinfo {author} {\bibfnamefont {K.}~\bibnamefont
  {G{\'{o}}ral}}, \ and\ \bibinfo {author} {\bibfnamefont {P.~S.}\ \bibnamefont
  {Julienne}},\ }\href {\doibase 10.1103/revmodphys.78.1311} {\bibfield
  {journal} {\bibinfo  {journal} {Rev. Mod. Phys.}\ }\textbf {\bibinfo {volume}
  {78}},\ \bibinfo {pages} {1311} (\bibinfo {year} {2006})}\BibitemShut
  {NoStop}%
\bibitem [{\citenamefont {Ferlaino}\ \emph {et~al.}(2009)\citenamefont
  {Ferlaino}, \citenamefont {Knoop},\ and\ \citenamefont
  {Grimm}}]{Ferlaino:2009}%
  \BibitemOpen
  \bibfield  {author} {\bibinfo {author} {\bibfnamefont {F.}~\bibnamefont
  {Ferlaino}}, \bibinfo {author} {\bibfnamefont {S.}~\bibnamefont {Knoop}}, \
  and\ \bibinfo {author} {\bibfnamefont {R.}~\bibnamefont {Grimm}},\ }\href
  {\doibase 10.1201/9781420059045} {\emph {\bibinfo {title} {Cold
  Molecules}}},\ edited by\ \bibinfo {editor} {\bibfnamefont {R.}~\bibnamefont
  {Krems}}, \bibinfo {editor} {\bibfnamefont {B.}~\bibnamefont {Friedrich}}, \
  and\ \bibinfo {editor} {\bibfnamefont {W.~C.}\ \bibnamefont {Stwalley}}\
  (\bibinfo  {publisher} {{CRC} Press},\ \bibinfo {year} {2009})\ Chap.\
  \bibinfo {chapter} {9, Ultracold Feshbach Molecules}\BibitemShut {NoStop}%
\bibitem [{\citenamefont {Chin}\ \emph
  {et~al.}(2010{\natexlab{a}})\citenamefont {Chin}, \citenamefont {Grimm},
  \citenamefont {Julienne},\ and\ \citenamefont {Tiesinga}}]{Julienne:2010}%
  \BibitemOpen
  \bibfield  {author} {\bibinfo {author} {\bibfnamefont {C.}~\bibnamefont
  {Chin}}, \bibinfo {author} {\bibfnamefont {R.}~\bibnamefont {Grimm}},
  \bibinfo {author} {\bibfnamefont {P.~S.}\ \bibnamefont {Julienne}}, \ and\
  \bibinfo {author} {\bibfnamefont {E.}~\bibnamefont {Tiesinga}},\ }\href
  {\doibase 10.1103/RevModPhys.82.1225} {\bibfield  {journal} {\bibinfo
  {journal} {Rev. Mod. Phys.}\ }\textbf {\bibinfo {volume} {82}},\ \bibinfo
  {pages} {1225} (\bibinfo {year} {2010}{\natexlab{a}})}\BibitemShut {NoStop}%
\bibitem [{\citenamefont {Tomza}\ \emph {et~al.}(2019)\citenamefont {Tomza},
  \citenamefont {Jachymski}, \citenamefont {Gerritsma}, \citenamefont
  {Negretti}, \citenamefont {Calarco}, \citenamefont {Idziaszek},\ and\
  \citenamefont {Julienne}}]{Tomza:2019}%
  \BibitemOpen
  \bibfield  {author} {\bibinfo {author} {\bibfnamefont {M.}~\bibnamefont
  {Tomza}}, \bibinfo {author} {\bibfnamefont {K.}~\bibnamefont {Jachymski}},
  \bibinfo {author} {\bibfnamefont {R.}~\bibnamefont {Gerritsma}}, \bibinfo
  {author} {\bibfnamefont {A.}~\bibnamefont {Negretti}}, \bibinfo {author}
  {\bibfnamefont {T.}~\bibnamefont {Calarco}}, \bibinfo {author} {\bibfnamefont
  {Z.}~\bibnamefont {Idziaszek}}, \ and\ \bibinfo {author} {\bibfnamefont
  {P.~S.}\ \bibnamefont {Julienne}},\ }\href {\doibase
  10.1103/revmodphys.91.035001} {\bibfield  {journal} {\bibinfo  {journal}
  {Rev. Mod. Phys.}\ }\textbf {\bibinfo {volume} {91}},\ \bibinfo {pages}
  {035001} (\bibinfo {year} {2019})}\BibitemShut {NoStop}%
\bibitem [{\citenamefont {Ratschbacher}\ \emph {et~al.}(2012)\citenamefont
  {Ratschbacher}, \citenamefont {Zipkes}, \citenamefont {Sias},\ and\
  \citenamefont {Köhl}}]{Ratschbacher:2012}%
  \BibitemOpen
  \bibfield  {author} {\bibinfo {author} {\bibfnamefont {L.}~\bibnamefont
  {Ratschbacher}}, \bibinfo {author} {\bibfnamefont {C.}~\bibnamefont
  {Zipkes}}, \bibinfo {author} {\bibfnamefont {C.}~\bibnamefont {Sias}}, \ and\
  \bibinfo {author} {\bibfnamefont {M.}~\bibnamefont {Köhl}},\ }\href
  {\doibase 10.1038/nphys2373} {\bibfield  {journal} {\bibinfo  {journal} {Nat.
  Phys.}\ }\textbf {\bibinfo {volume} {8}},\ \bibinfo {pages} {649} (\bibinfo
  {year} {2012})}\BibitemShut {NoStop}%
\bibitem [{\citenamefont {Haze}\ \emph {et~al.}(2015)\citenamefont {Haze},
  \citenamefont {Saito}, \citenamefont {Fujinaga},\ and\ \citenamefont
  {Mukaiyama}}]{Haze:2015}%
  \BibitemOpen
  \bibfield  {author} {\bibinfo {author} {\bibfnamefont {S.}~\bibnamefont
  {Haze}}, \bibinfo {author} {\bibfnamefont {R.}~\bibnamefont {Saito}},
  \bibinfo {author} {\bibfnamefont {M.}~\bibnamefont {Fujinaga}}, \ and\
  \bibinfo {author} {\bibfnamefont {T.}~\bibnamefont {Mukaiyama}},\ }\href
  {\doibase 10.1103/PhysRevA.91.032709} {\bibfield  {journal} {\bibinfo
  {journal} {Phys. Rev. A}\ }\textbf {\bibinfo {volume} {91}},\ \bibinfo
  {pages} {032709} (\bibinfo {year} {2015})}\BibitemShut {NoStop}%
\bibitem [{\citenamefont {Jyothi}\ \emph {et~al.}(2016)\citenamefont {Jyothi},
  \citenamefont {Ray}, \citenamefont {Dutta}, \citenamefont {Allouche},
  \citenamefont {Vexiau}, \citenamefont {Dulieu},\ and\ \citenamefont
  {Rangwala}}]{Jyothi2016}%
  \BibitemOpen
  \bibfield  {author} {\bibinfo {author} {\bibfnamefont {S.}~\bibnamefont
  {Jyothi}}, \bibinfo {author} {\bibfnamefont {T.}~\bibnamefont {Ray}},
  \bibinfo {author} {\bibfnamefont {S.}~\bibnamefont {Dutta}}, \bibinfo
  {author} {\bibfnamefont {A.}~\bibnamefont {Allouche}}, \bibinfo {author}
  {\bibfnamefont {R.}~\bibnamefont {Vexiau}}, \bibinfo {author} {\bibfnamefont
  {O.}~\bibnamefont {Dulieu}}, \ and\ \bibinfo {author} {\bibfnamefont
  {S.}~\bibnamefont {Rangwala}},\ }\href {\doibase
  10.1103/physrevlett.117.213002} {\bibfield  {journal} {\bibinfo  {journal}
  {Phys. Rev. Lett.}\ }\textbf {\bibinfo {volume} {117}},\ \bibinfo {pages}
  {213002} (\bibinfo {year} {2016})}\BibitemShut {NoStop}%
\bibitem [{\citenamefont {Joger}\ \emph {et~al.}(2017)\citenamefont {Joger},
  \citenamefont {F\"urst}, \citenamefont {Ewald}, \citenamefont {Feldker},
  \citenamefont {Tomza},\ and\ \citenamefont {Gerritsma}}]{Joger:2017}%
  \BibitemOpen
  \bibfield  {author} {\bibinfo {author} {\bibfnamefont {J.}~\bibnamefont
  {Joger}}, \bibinfo {author} {\bibfnamefont {H.}~\bibnamefont {F\"urst}},
  \bibinfo {author} {\bibfnamefont {N.}~\bibnamefont {Ewald}}, \bibinfo
  {author} {\bibfnamefont {T.}~\bibnamefont {Feldker}}, \bibinfo {author}
  {\bibfnamefont {M.}~\bibnamefont {Tomza}}, \ and\ \bibinfo {author}
  {\bibfnamefont {R.}~\bibnamefont {Gerritsma}},\ }\href {\doibase
  10.1103/PhysRevA.96.030703} {\bibfield  {journal} {\bibinfo  {journal} {Phys.
  Rev. A}\ }\textbf {\bibinfo {volume} {96}},\ \bibinfo {pages} {030703(R)}
  (\bibinfo {year} {2017})}\BibitemShut {NoStop}%
\bibitem [{\citenamefont {Ratschbacher}\ \emph {et~al.}(2013)\citenamefont
  {Ratschbacher}, \citenamefont {Sias}, \citenamefont {Carcagni}, \citenamefont
  {Silver}, \citenamefont {Zipkes},\ and\ \citenamefont
  {K\"ohl}}]{Ratschbacher:2013}%
  \BibitemOpen
  \bibfield  {author} {\bibinfo {author} {\bibfnamefont {L.}~\bibnamefont
  {Ratschbacher}}, \bibinfo {author} {\bibfnamefont {C.}~\bibnamefont {Sias}},
  \bibinfo {author} {\bibfnamefont {L.}~\bibnamefont {Carcagni}}, \bibinfo
  {author} {\bibfnamefont {J.~M.}\ \bibnamefont {Silver}}, \bibinfo {author}
  {\bibfnamefont {C.}~\bibnamefont {Zipkes}}, \ and\ \bibinfo {author}
  {\bibfnamefont {M.}~\bibnamefont {K\"ohl}},\ }\href {\doibase
  10.1103/PhysRevLett.110.160402} {\bibfield  {journal} {\bibinfo  {journal}
  {Phys.~Rev.~Lett.}\ }\textbf {\bibinfo {volume} {110}},\ \bibinfo {pages}
  {160402} (\bibinfo {year} {2013})}\BibitemShut {NoStop}%
\bibitem [{\citenamefont {Sikorsky}\ \emph {et~al.}(2018)\citenamefont
  {Sikorsky}, \citenamefont {Meir}, \citenamefont {Ben-shlomi}, \citenamefont
  {Akerman},\ and\ \citenamefont {Ozeri}}]{Sikorsky:2018}%
  \BibitemOpen
  \bibfield  {author} {\bibinfo {author} {\bibfnamefont {T.}~\bibnamefont
  {Sikorsky}}, \bibinfo {author} {\bibfnamefont {Z.}~\bibnamefont {Meir}},
  \bibinfo {author} {\bibfnamefont {R.}~\bibnamefont {Ben-shlomi}}, \bibinfo
  {author} {\bibfnamefont {N.}~\bibnamefont {Akerman}}, \ and\ \bibinfo
  {author} {\bibfnamefont {R.}~\bibnamefont {Ozeri}},\ }\href {\doibase
  10.1038/s41467-018-03373-y} {\bibfield  {journal} {\bibinfo  {journal} {Nat.
  Comm.}\ }\textbf {\bibinfo {volume} {9}},\ \bibinfo {pages} {920} (\bibinfo
  {year} {2018})}\BibitemShut {NoStop}%
\bibitem [{\citenamefont {F\"urst}\ \emph {et~al.}(2018)\citenamefont
  {F\"urst}, \citenamefont {Feldker}, \citenamefont {Ewald}, \citenamefont
  {Joger}, \citenamefont {Tomza},\ and\ \citenamefont
  {Gerritsma}}]{Fuerst:2018:spin}%
  \BibitemOpen
  \bibfield  {author} {\bibinfo {author} {\bibfnamefont {H.}~\bibnamefont
  {F\"urst}}, \bibinfo {author} {\bibfnamefont {T.}~\bibnamefont {Feldker}},
  \bibinfo {author} {\bibfnamefont {N.~V.}\ \bibnamefont {Ewald}}, \bibinfo
  {author} {\bibfnamefont {J.}~\bibnamefont {Joger}}, \bibinfo {author}
  {\bibfnamefont {M.}~\bibnamefont {Tomza}}, \ and\ \bibinfo {author}
  {\bibfnamefont {R.}~\bibnamefont {Gerritsma}},\ }\href {\doibase
  10.1103/PhysRevA.98.012713} {\bibfield  {journal} {\bibinfo  {journal} {Phys.
  Rev. A}\ }\textbf {\bibinfo {volume} {98}},\ \bibinfo {pages} {012713}
  (\bibinfo {year} {2018})}\BibitemShut {NoStop}%
\bibitem [{\citenamefont {Härter}\ \emph {et~al.}(2012)\citenamefont
  {Härter}, \citenamefont {Krükow}, \citenamefont {Brunner}, \citenamefont
  {Schnitzler}, \citenamefont {Schmid},\ and\ \citenamefont
  {Denschlag}}]{Haerter2012}%
  \BibitemOpen
  \bibfield  {author} {\bibinfo {author} {\bibfnamefont {A.}~\bibnamefont
  {Härter}}, \bibinfo {author} {\bibfnamefont {A.}~\bibnamefont {Krükow}},
  \bibinfo {author} {\bibfnamefont {A.}~\bibnamefont {Brunner}}, \bibinfo
  {author} {\bibfnamefont {W.}~\bibnamefont {Schnitzler}}, \bibinfo {author}
  {\bibfnamefont {S.}~\bibnamefont {Schmid}}, \ and\ \bibinfo {author}
  {\bibfnamefont {J.~H.}\ \bibnamefont {Denschlag}},\ }\href {\doibase
  10.1103/physrevlett.109.123201} {\bibfield  {journal} {\bibinfo  {journal}
  {Phys.~Rev.~Lett.}\ }\textbf {\bibinfo {volume} {109}},\ \bibinfo {pages}
  {123201} (\bibinfo {year} {2012})}\BibitemShut {NoStop}%
\bibitem [{\citenamefont {P{\'{e}}rez-R{\'{\i}}os}\ and\ \citenamefont
  {Greene}(2015)}]{PerezRios2015}%
  \BibitemOpen
  \bibfield  {author} {\bibinfo {author} {\bibfnamefont {J.}~\bibnamefont
  {P{\'{e}}rez-R{\'{\i}}os}}\ and\ \bibinfo {author} {\bibfnamefont {C.~H.}\
  \bibnamefont {Greene}},\ }\href {\doibase 10.1063/1.4927702} {\bibfield
  {journal} {\bibinfo  {journal} {J. Chem. Phys.}\ }\textbf {\bibinfo {volume}
  {143}},\ \bibinfo {pages} {041105} (\bibinfo {year} {2015})}\BibitemShut
  {NoStop}%
\bibitem [{\citenamefont {Krükow}\ \emph {et~al.}(2016)\citenamefont
  {Krükow}, \citenamefont {Mohammadi}, \citenamefont {Härter},\ and\
  \citenamefont {Denschlag}}]{Kruekow2016}%
  \BibitemOpen
  \bibfield  {author} {\bibinfo {author} {\bibfnamefont {A.}~\bibnamefont
  {Krükow}}, \bibinfo {author} {\bibfnamefont {A.}~\bibnamefont {Mohammadi}},
  \bibinfo {author} {\bibfnamefont {A.}~\bibnamefont {Härter}}, \ and\
  \bibinfo {author} {\bibfnamefont {J.~H.}\ \bibnamefont {Denschlag}},\ }\href
  {https://journals.aps.org/pra/abstract/10.1103/PhysRevA.94.030701} {\bibfield
   {journal} {\bibinfo  {journal} {Phys. Rev. A}\ }\textbf {\bibinfo {volume}
  {94}},\ \bibinfo {pages} {030701(R)} (\bibinfo {year} {2016})}\BibitemShut
  {NoStop}%
\bibitem [{\citenamefont {Mohammadi}\ \emph {et~al.}(2021)\citenamefont
  {Mohammadi}, \citenamefont {Krükow}, \citenamefont {Mahdian}, \citenamefont
  {Dei{\ss}}, \citenamefont {P{\'{e}}rez-R{\'{\i}}os}, \citenamefont
  {da~Silva}, \citenamefont {Raoult}, \citenamefont {Dulieu},\ and\
  \citenamefont {Denschlag}}]{Mohammadi2021}%
  \BibitemOpen
  \bibfield  {author} {\bibinfo {author} {\bibfnamefont {A.}~\bibnamefont
  {Mohammadi}}, \bibinfo {author} {\bibfnamefont {A.}~\bibnamefont {Krükow}},
  \bibinfo {author} {\bibfnamefont {A.}~\bibnamefont {Mahdian}}, \bibinfo
  {author} {\bibfnamefont {M.}~\bibnamefont {Dei{\ss}}}, \bibinfo {author}
  {\bibfnamefont {J.}~\bibnamefont {P{\'{e}}rez-R{\'{\i}}os}}, \bibinfo
  {author} {\bibfnamefont {H.}~\bibnamefont {da~Silva}}, \bibinfo {author}
  {\bibfnamefont {M.}~\bibnamefont {Raoult}}, \bibinfo {author} {\bibfnamefont
  {O.}~\bibnamefont {Dulieu}}, \ and\ \bibinfo {author} {\bibfnamefont {J.~H.}\
  \bibnamefont {Denschlag}},\ }\href
  {https://journals.aps.org/prresearch/abstract/10.1103/PhysRevResearch.3.013196}
  {\bibfield  {journal} {\bibinfo  {journal} {Phys. Rev. Research}\ }\textbf
  {\bibinfo {volume} {3}},\ \bibinfo {pages} {013196} (\bibinfo {year}
  {2021})}\BibitemShut {NoStop}%
\bibitem [{\citenamefont {Monroe}\ \emph {et~al.}(1995)\citenamefont {Monroe},
  \citenamefont {Meekhof}, \citenamefont {King}, \citenamefont {Itano},\ and\
  \citenamefont {Wineland}}]{Monroe:1995}%
  \BibitemOpen
  \bibfield  {author} {\bibinfo {author} {\bibfnamefont {C.}~\bibnamefont
  {Monroe}}, \bibinfo {author} {\bibfnamefont {D.~M.}\ \bibnamefont {Meekhof}},
  \bibinfo {author} {\bibfnamefont {B.~E.}\ \bibnamefont {King}}, \bibinfo
  {author} {\bibfnamefont {W.~M.}\ \bibnamefont {Itano}}, \ and\ \bibinfo
  {author} {\bibfnamefont {D.~J.}\ \bibnamefont {Wineland}},\ }\href {\doibase
  10.1103/physrevlett.75.4714} {\bibfield  {journal} {\bibinfo  {journal}
  {Phys. Rev. Lett.}\ }\textbf {\bibinfo {volume} {75}},\ \bibinfo {pages}
  {4714} (\bibinfo {year} {1995})}\BibitemShut {NoStop}%
\bibitem [{\citenamefont {Leibfried}\ \emph {et~al.}(2003)\citenamefont
  {Leibfried}, \citenamefont {Blatt}, \citenamefont {Monroe},\ and\
  \citenamefont {Wineland}}]{Leibfried:2003}%
  \BibitemOpen
  \bibfield  {author} {\bibinfo {author} {\bibfnamefont {D.}~\bibnamefont
  {Leibfried}}, \bibinfo {author} {\bibfnamefont {R.}~\bibnamefont {Blatt}},
  \bibinfo {author} {\bibfnamefont {C.}~\bibnamefont {Monroe}}, \ and\ \bibinfo
  {author} {\bibfnamefont {D.}~\bibnamefont {Wineland}},\ }\href {\doibase
  10.1103/RevModPhys.75.281} {\bibfield  {journal} {\bibinfo  {journal}
  {Rev.~Mod.~Phys.}\ }\textbf {\bibinfo {volume} {75}},\ \bibinfo {pages} {281}
  (\bibinfo {year} {2003})}\BibitemShut {NoStop}%
\bibitem [{\citenamefont {Schmid}\ \emph {et~al.}(2010)\citenamefont {Schmid},
  \citenamefont {H\"arter},\ and\ \citenamefont {Denschlag}}]{Schmid:2010}%
  \BibitemOpen
  \bibfield  {author} {\bibinfo {author} {\bibfnamefont {S.}~\bibnamefont
  {Schmid}}, \bibinfo {author} {\bibfnamefont {A.}~\bibnamefont {H\"arter}}, \
  and\ \bibinfo {author} {\bibfnamefont {J.~H.}\ \bibnamefont {Denschlag}},\
  }\href {\doibase 10.1103/PhysRevLett.105.133202} {\bibfield  {journal}
  {\bibinfo  {journal} {Phys. Rev. Lett.}\ }\textbf {\bibinfo {volume} {105}},\
  \bibinfo {pages} {133202} (\bibinfo {year} {2010})}\BibitemShut {NoStop}%
\bibitem [{\citenamefont {Zwerger}(2012)}]{Zwerger2012tbb}%
  \BibitemOpen
  \bibinfo {editor} {\bibfnamefont {W.}~\bibnamefont {Zwerger}},\ ed.,\
  \href@noop {} {\emph {\bibinfo {title}
  {\href{http://www.springer.com/de/book/9783642219771}{The BCS-BEC Crossover
  and the Unitary Fermi Gas}}}},\ \bibinfo {series} {Lecture Notes in Physics},
  Vol.\ \bibinfo {volume} {836}\ (\bibinfo  {publisher} {Springer, Berlin
  Heidelberg},\ \bibinfo {year} {2012})\BibitemShut {NoStop}%
\bibitem [{\citenamefont {Astrakharchik}\ \emph {et~al.}(2021)\citenamefont
  {Astrakharchik}, \citenamefont {Ardila}, \citenamefont {Schmidt},
  \citenamefont {Jachymski},\ and\ \citenamefont
  {Negretti}}]{Astrakharchik:2021}%
  \BibitemOpen
  \bibfield  {author} {\bibinfo {author} {\bibfnamefont {G.~E.}\ \bibnamefont
  {Astrakharchik}}, \bibinfo {author} {\bibfnamefont {L.~A.~P.}\ \bibnamefont
  {Ardila}}, \bibinfo {author} {\bibfnamefont {R.}~\bibnamefont {Schmidt}},
  \bibinfo {author} {\bibfnamefont {K.}~\bibnamefont {Jachymski}}, \ and\
  \bibinfo {author} {\bibfnamefont {A.}~\bibnamefont {Negretti}},\ }\href
  {https://www.nature.com/articles/s42005-021-00597-1} {\bibfield  {journal}
  {\bibinfo  {journal} {Comm. Phys.}\ }\textbf {\bibinfo {volume} {4}},\
  \bibinfo {pages} {94} (\bibinfo {year} {2021})}\BibitemShut {NoStop}%
\bibitem [{\citenamefont {Christensen}\ \emph {et~al.}(2021)\citenamefont
  {Christensen}, \citenamefont {Camacho-Guardian},\ and\ \citenamefont
  {Bruun}}]{Christensen2021cpa}%
  \BibitemOpen
  \bibfield  {author} {\bibinfo {author} {\bibfnamefont {E.~R.}\ \bibnamefont
  {Christensen}}, \bibinfo {author} {\bibfnamefont {A.}~\bibnamefont
  {Camacho-Guardian}}, \ and\ \bibinfo {author} {\bibfnamefont {G.~M.}\
  \bibnamefont {Bruun}},\ }\href {\doibase 10.1103/PhysRevLett.126.243001}
  {\bibfield  {journal} {\bibinfo  {journal} {Phys. Rev. Lett.}\ }\textbf
  {\bibinfo {volume} {126}},\ \bibinfo {pages} {243001} (\bibinfo {year}
  {2021})}\BibitemShut {NoStop}%
\bibitem [{\citenamefont {Oghittu}\ \emph {et~al.}(2021)\citenamefont
  {Oghittu}, \citenamefont {Johannsen}, \citenamefont {Negretti},\ and\
  \citenamefont {Gerritsma}}]{Oghittu2021a}%
  \BibitemOpen
  \bibfield  {author} {\bibinfo {author} {\bibfnamefont {L.}~\bibnamefont
  {Oghittu}}, \bibinfo {author} {\bibfnamefont {M.}~\bibnamefont {Johannsen}},
  \bibinfo {author} {\bibfnamefont {A.}~\bibnamefont {Negretti}}, \ and\
  \bibinfo {author} {\bibfnamefont {R.}~\bibnamefont {Gerritsma}},\ }\href
  {\doibase 10.1103/physreva.104.053314} {\bibfield  {journal} {\bibinfo
  {journal} {Phys. Rev. A}\ }\textbf {\bibinfo {volume} {104}},\ \bibinfo
  {pages} {053314} (\bibinfo {year} {2021})}\BibitemShut {NoStop}%
\bibitem [{\citenamefont {Mur-Petit}\ \emph {et~al.}(2012)\citenamefont
  {Mur-Petit}, \citenamefont {Garc{\'{\i}}a-Ripoll}, \citenamefont
  {P{\'{e}}rez-R{\'{\i}}os}, \citenamefont {Campos-Mart{\'{\i}}nez},
  \citenamefont {Hern{\'{a}}ndez},\ and\ \citenamefont
  {Willitsch}}]{MurPetit:2012}%
  \BibitemOpen
  \bibfield  {author} {\bibinfo {author} {\bibfnamefont {J.}~\bibnamefont
  {Mur-Petit}}, \bibinfo {author} {\bibfnamefont {J.~J.}\ \bibnamefont
  {Garc{\'{\i}}a-Ripoll}}, \bibinfo {author} {\bibfnamefont {J.}~\bibnamefont
  {P{\'{e}}rez-R{\'{\i}}os}}, \bibinfo {author} {\bibfnamefont
  {J.}~\bibnamefont {Campos-Mart{\'{\i}}nez}}, \bibinfo {author} {\bibfnamefont
  {M.~I.}\ \bibnamefont {Hern{\'{a}}ndez}}, \ and\ \bibinfo {author}
  {\bibfnamefont {S.}~\bibnamefont {Willitsch}},\ }\href {\doibase
  10.1103/physreva.85.022308} {\bibfield  {journal} {\bibinfo  {journal}
  {Phys.~Rev.~A}\ }\textbf {\bibinfo {volume} {85}},\ \bibinfo {pages} {022308}
  (\bibinfo {year} {2012})}\BibitemShut {NoStop}%
\bibitem [{\citenamefont {Khanyile}\ \emph {et~al.}(2015)\citenamefont
  {Khanyile}, \citenamefont {Shu},\ and\ \citenamefont {Brown}}]{Khanyile2015}%
  \BibitemOpen
  \bibfield  {author} {\bibinfo {author} {\bibfnamefont {N.~B.}\ \bibnamefont
  {Khanyile}}, \bibinfo {author} {\bibfnamefont {G.}~\bibnamefont {Shu}}, \
  and\ \bibinfo {author} {\bibfnamefont {K.~R.}\ \bibnamefont {Brown}},\ }\href
  {https://www.nature.com/articles/ncomms8825} {\bibfield  {journal} {\bibinfo
  {journal} {Nat. Comm.}\ }\textbf {\bibinfo {volume} {6}},\ \bibinfo {pages}
  {7825} (\bibinfo {year} {2015})}\BibitemShut {NoStop}%
\bibitem [{\citenamefont {Wolf}\ \emph {et~al.}(2016)\citenamefont {Wolf},
  \citenamefont {Wan}, \citenamefont {Heip}, \citenamefont {Gebert},
  \citenamefont {Shi},\ and\ \citenamefont {Schmidt}}]{Wolf:2016}%
  \BibitemOpen
  \bibfield  {author} {\bibinfo {author} {\bibfnamefont {F.}~\bibnamefont
  {Wolf}}, \bibinfo {author} {\bibfnamefont {Y.}~\bibnamefont {Wan}}, \bibinfo
  {author} {\bibfnamefont {J.~C.}\ \bibnamefont {Heip}}, \bibinfo {author}
  {\bibfnamefont {F.}~\bibnamefont {Gebert}}, \bibinfo {author} {\bibfnamefont
  {C.}~\bibnamefont {Shi}}, \ and\ \bibinfo {author} {\bibfnamefont {P.~O.}\
  \bibnamefont {Schmidt}},\ }\href {\doibase 10.1038/nature16513} {\bibfield
  {journal} {\bibinfo  {journal} {Nature}\ }\textbf {\bibinfo {volume} {530}},\
  \bibinfo {pages} {457} (\bibinfo {year} {2016})}\BibitemShut {NoStop}%
\bibitem [{\citenamefont {wan Chou}\ \emph {et~al.}(2017)\citenamefont {wan
  Chou}, \citenamefont {Kurz}, \citenamefont {Hume}, \citenamefont {Plessow},
  \citenamefont {Leibrandt},\ and\ \citenamefont {Leibfried}}]{Chou:2017}%
  \BibitemOpen
  \bibfield  {author} {\bibinfo {author} {\bibfnamefont {C.}~\bibnamefont {wan
  Chou}}, \bibinfo {author} {\bibfnamefont {C.}~\bibnamefont {Kurz}}, \bibinfo
  {author} {\bibfnamefont {D.~B.}\ \bibnamefont {Hume}}, \bibinfo {author}
  {\bibfnamefont {P.~N.}\ \bibnamefont {Plessow}}, \bibinfo {author}
  {\bibfnamefont {D.~R.}\ \bibnamefont {Leibrandt}}, \ and\ \bibinfo {author}
  {\bibfnamefont {D.}~\bibnamefont {Leibfried}},\ }\href {\doibase
  10.1038/nature22338} {\bibfield  {journal} {\bibinfo  {journal} {Nature}\
  }\textbf {\bibinfo {volume} {545}},\ \bibinfo {pages} {203} (\bibinfo {year}
  {2017})}\BibitemShut {NoStop}%
\bibitem [{\citenamefont {Sinhal}\ \emph {et~al.}(2020)\citenamefont {Sinhal},
  \citenamefont {Meir}, \citenamefont {Najafian}, \citenamefont {Hegi},\ and\
  \citenamefont {Willitsch}}]{Sinhal:2020}%
  \BibitemOpen
  \bibfield  {author} {\bibinfo {author} {\bibfnamefont {M.}~\bibnamefont
  {Sinhal}}, \bibinfo {author} {\bibfnamefont {Z.}~\bibnamefont {Meir}},
  \bibinfo {author} {\bibfnamefont {K.}~\bibnamefont {Najafian}}, \bibinfo
  {author} {\bibfnamefont {G.}~\bibnamefont {Hegi}}, \ and\ \bibinfo {author}
  {\bibfnamefont {S.}~\bibnamefont {Willitsch}},\ }\href {\doibase
  10.1126/science.aaz9837} {\bibfield  {journal} {\bibinfo  {journal}
  {Science}\ }\textbf {\bibinfo {volume} {367}},\ \bibinfo {pages} {1213}
  (\bibinfo {year} {2020})}\BibitemShut {NoStop}%
\bibitem [{\citenamefont {Katz}\ \emph {et~al.}(2021)\citenamefont {Katz},
  \citenamefont {Pinkas}, \citenamefont {Akerman},\ and\ \citenamefont
  {Ozeri}}]{Katz2021}%
  \BibitemOpen
  \bibfield  {author} {\bibinfo {author} {\bibfnamefont {O.}~\bibnamefont
  {Katz}}, \bibinfo {author} {\bibfnamefont {M.}~\bibnamefont {Pinkas}},
  \bibinfo {author} {\bibfnamefont {N.}~\bibnamefont {Akerman}}, \ and\
  \bibinfo {author} {\bibfnamefont {R.}~\bibnamefont {Ozeri}},\ }\href@noop {}
  {\  (\bibinfo {year} {2021})},\ \Eprint {http://arxiv.org/abs/2107.08441}
  {arXiv:2107.08441 [physics.atom-ph]} \BibitemShut {NoStop}%
\bibitem [{\citenamefont {Hirzler}\ \emph
  {et~al.}(2020{\natexlab{a}})\citenamefont {Hirzler}, \citenamefont {Feldker},
  \citenamefont {Fürst}, \citenamefont {Ewald}, \citenamefont {Trimby},
  \citenamefont {Lous}, \citenamefont {Espinoza}, \citenamefont {Mazzanti},
  \citenamefont {Joger},\ and\ \citenamefont {Gerritsma}}]{Hirzler2020}%
  \BibitemOpen
  \bibfield  {author} {\bibinfo {author} {\bibfnamefont {H.}~\bibnamefont
  {Hirzler}}, \bibinfo {author} {\bibfnamefont {T.}~\bibnamefont {Feldker}},
  \bibinfo {author} {\bibfnamefont {H.}~\bibnamefont {Fürst}}, \bibinfo
  {author} {\bibfnamefont {N.~V.}\ \bibnamefont {Ewald}}, \bibinfo {author}
  {\bibfnamefont {E.}~\bibnamefont {Trimby}}, \bibinfo {author} {\bibfnamefont
  {R.~S.}\ \bibnamefont {Lous}}, \bibinfo {author} {\bibfnamefont {J.~D.~A.}\
  \bibnamefont {Espinoza}}, \bibinfo {author} {\bibfnamefont {M.}~\bibnamefont
  {Mazzanti}}, \bibinfo {author} {\bibfnamefont {J.}~\bibnamefont {Joger}}, \
  and\ \bibinfo {author} {\bibfnamefont {R.}~\bibnamefont {Gerritsma}},\ }\href
  {https://journals.aps.org/pra/abstract/10.1103/PhysRevA.102.033109}
  {\bibfield  {journal} {\bibinfo  {journal} {Phys. Rev. A}\ }\textbf {\bibinfo
  {volume} {102}},\ \bibinfo {pages} {033109} (\bibinfo {year}
  {2020}{\natexlab{a}})}\BibitemShut {NoStop}%
\bibitem [{\citenamefont {Z\"urn}\ \emph {et~al.}(2013)\citenamefont {Z\"urn},
  \citenamefont {Lompe}, \citenamefont {Wenz}, \citenamefont {Jochim},
  \citenamefont {Julienne},\ and\ \citenamefont {Hutson}}]{Zuern:2013}%
  \BibitemOpen
  \bibfield  {author} {\bibinfo {author} {\bibfnamefont {G.}~\bibnamefont
  {Z\"urn}}, \bibinfo {author} {\bibfnamefont {T.}~\bibnamefont {Lompe}},
  \bibinfo {author} {\bibfnamefont {A.~N.}\ \bibnamefont {Wenz}}, \bibinfo
  {author} {\bibfnamefont {S.}~\bibnamefont {Jochim}}, \bibinfo {author}
  {\bibfnamefont {P.~S.}\ \bibnamefont {Julienne}}, \ and\ \bibinfo {author}
  {\bibfnamefont {J.~M.}\ \bibnamefont {Hutson}},\ }\href {\doibase
  10.1103/physrevlett.110.135301} {\bibfield  {journal} {\bibinfo  {journal}
  {Phys.~Rev.~Lett.}\ }\textbf {\bibinfo {volume} {110}},\ \bibinfo {pages}
  {135301} (\bibinfo {year} {2013})}\BibitemShut {NoStop}%
\bibitem [{\citenamefont {Jochim}\ \emph {et~al.}(2003)\citenamefont {Jochim},
  \citenamefont {Bartenstein}, \citenamefont {Altmeyer}, \citenamefont {Hendl},
  \citenamefont {Riedl}, \citenamefont {Chin}, \citenamefont
  {Hecker~Denschlag},\ and\ \citenamefont {Grimm}}]{Jochim:2003}%
  \BibitemOpen
  \bibfield  {author} {\bibinfo {author} {\bibfnamefont {S.}~\bibnamefont
  {Jochim}}, \bibinfo {author} {\bibfnamefont {M.}~\bibnamefont {Bartenstein}},
  \bibinfo {author} {\bibfnamefont {A.}~\bibnamefont {Altmeyer}}, \bibinfo
  {author} {\bibfnamefont {G.}~\bibnamefont {Hendl}}, \bibinfo {author}
  {\bibfnamefont {S.}~\bibnamefont {Riedl}}, \bibinfo {author} {\bibfnamefont
  {C.}~\bibnamefont {Chin}}, \bibinfo {author} {\bibfnamefont {J.}~\bibnamefont
  {Hecker~Denschlag}}, \ and\ \bibinfo {author} {\bibfnamefont
  {R.}~\bibnamefont {Grimm}},\ }\href {\doibase 10.1126/science.1093280}
  {\bibfield  {journal} {\bibinfo  {journal} {Science}\ }\textbf {\bibinfo
  {volume} {302}},\ \bibinfo {pages} {2101} (\bibinfo {year}
  {2003})}\BibitemShut {NoStop}%
\bibitem [{\citenamefont {Chin}\ and\ \citenamefont {Grimm}(2004)}]{Chin2004}%
  \BibitemOpen
  \bibfield  {author} {\bibinfo {author} {\bibfnamefont {C.}~\bibnamefont
  {Chin}}\ and\ \bibinfo {author} {\bibfnamefont {R.}~\bibnamefont {Grimm}},\
  }\href {https://link.aps.org/doi/10.1103/PhysRevA.69.033612} {\bibfield
  {journal} {\bibinfo  {journal} {Phys. Rev. A}\ }\textbf {\bibinfo {volume}
  {69}},\ \bibinfo {pages} {033612} (\bibinfo {year} {2004})}\BibitemShut
  {NoStop}%
\bibitem [{Sup()}]{Suppl:2021}%
  \BibitemOpen
  \href@noop {} {\bibinfo  {journal} {See supplemental material at [URL will be
  inserted by publisher] for details, Refs. [10,11,22,24,36,38,40,41,44-52]}\
  }\BibitemShut {NoStop}%
\bibitem [{\citenamefont {Hirzler}\ \emph
  {et~al.}(2020{\natexlab{b}})\citenamefont {Hirzler}, \citenamefont {Trimby},
  \citenamefont {Lous}, \citenamefont {Groenenboom}, \citenamefont
  {Gerritsma},\ and\ \citenamefont {P{\'{e}}rez-R{\'{\i}}os}}]{Hirzler2020a}%
  \BibitemOpen
\bibfield  {journal} {  }\bibfield  {author} {\bibinfo {author} {\bibfnamefont
  {H.}~\bibnamefont {Hirzler}}, \bibinfo {author} {\bibfnamefont
  {E.}~\bibnamefont {Trimby}}, \bibinfo {author} {\bibfnamefont {R.~S.}\
  \bibnamefont {Lous}}, \bibinfo {author} {\bibfnamefont {G.~C.}\ \bibnamefont
  {Groenenboom}}, \bibinfo {author} {\bibfnamefont {R.}~\bibnamefont
  {Gerritsma}}, \ and\ \bibinfo {author} {\bibfnamefont {J.}~\bibnamefont
  {P{\'{e}}rez-R{\'{\i}}os}},\ }\href
  {https://journals.aps.org/prresearch/abstract/10.1103/PhysRevResearch.2.033232}
  {\bibfield  {journal} {\bibinfo  {journal} {Phys. Rev. Research}\ }\textbf
  {\bibinfo {volume} {2}},\ \bibinfo {pages} {033232} (\bibinfo {year}
  {2020}{\natexlab{b}})}\BibitemShut {NoStop}%
\bibitem [{\citenamefont {Weckesser}\ \emph {et~al.}(2021)\citenamefont
  {Weckesser}, \citenamefont {Thielemann}, \citenamefont {Wiater},
  \citenamefont {Wojciechowska}, \citenamefont {Karpa}, \citenamefont
  {Jachymski}, \citenamefont {Tomza}, \citenamefont {Walker},\ and\
  \citenamefont {Schätz}}]{Weckesser:2021}%
  \BibitemOpen
  \bibfield  {author} {\bibinfo {author} {\bibfnamefont {P.}~\bibnamefont
  {Weckesser}}, \bibinfo {author} {\bibfnamefont {F.}~\bibnamefont
  {Thielemann}}, \bibinfo {author} {\bibfnamefont {D.}~\bibnamefont {Wiater}},
  \bibinfo {author} {\bibfnamefont {A.}~\bibnamefont {Wojciechowska}}, \bibinfo
  {author} {\bibfnamefont {L.}~\bibnamefont {Karpa}}, \bibinfo {author}
  {\bibfnamefont {K.}~\bibnamefont {Jachymski}}, \bibinfo {author}
  {\bibfnamefont {M.}~\bibnamefont {Tomza}}, \bibinfo {author} {\bibfnamefont
  {T.}~\bibnamefont {Walker}}, \ and\ \bibinfo {author} {\bibfnamefont
  {T.}~\bibnamefont {Schätz}},\ }\href {https://arxiv.org/abs/2105.09382}
  {\bibfield  {journal} {\bibinfo  {journal} {arxiv:2105.09382}\ } (\bibinfo
  {year} {2021})}\BibitemShut {NoStop}%
\bibitem [{\citenamefont {Feldker}\ \emph {et~al.}(2020)\citenamefont
  {Feldker}, \citenamefont {F\"urst}, \citenamefont {Hirzler}, \citenamefont
  {Ewald}, \citenamefont {Mazzanti}, \citenamefont {Wiater}, \citenamefont
  {Tomza},\ and\ \citenamefont {Gerritsma}}]{Feldker:2020}%
  \BibitemOpen
  \bibfield  {author} {\bibinfo {author} {\bibfnamefont {T.}~\bibnamefont
  {Feldker}}, \bibinfo {author} {\bibfnamefont {H.}~\bibnamefont {F\"urst}},
  \bibinfo {author} {\bibfnamefont {H.}~\bibnamefont {Hirzler}}, \bibinfo
  {author} {\bibfnamefont {N.~V.}\ \bibnamefont {Ewald}}, \bibinfo {author}
  {\bibfnamefont {M.}~\bibnamefont {Mazzanti}}, \bibinfo {author}
  {\bibfnamefont {D.}~\bibnamefont {Wiater}}, \bibinfo {author} {\bibfnamefont
  {M.}~\bibnamefont {Tomza}}, \ and\ \bibinfo {author} {\bibfnamefont
  {R.}~\bibnamefont {Gerritsma}},\ }\href {\doibase 10.1038/s41567-019-0772-5}
  {\bibfield  {journal} {\bibinfo  {journal} {Nat. Phys.}\ }\textbf {\bibinfo
  {volume} {16}},\ \bibinfo {pages} {413} (\bibinfo {year} {2020})}\BibitemShut
  {NoStop}%
\bibitem [{Note1()}]{Note1}%
  \BibitemOpen
  \bibinfo {note} {This is also confirmed by separate time-of-flight
  measurements of the atom density for varying $B_\protect \mathrm
  {Li_2}$}\BibitemShut {NoStop}%
\bibitem [{\citenamefont {Kokkelmans}\ \emph {et~al.}(2004)\citenamefont
  {Kokkelmans}, \citenamefont {Shlyapnikov},\ and\ \citenamefont
  {Salomon}}]{Kokkelmans2004}%
  \BibitemOpen
  \bibfield  {author} {\bibinfo {author} {\bibfnamefont {S.~J. J. M.~F.}\
  \bibnamefont {Kokkelmans}}, \bibinfo {author} {\bibfnamefont {G.~V.}\
  \bibnamefont {Shlyapnikov}}, \ and\ \bibinfo {author} {\bibfnamefont
  {C.}~\bibnamefont {Salomon}},\ }\href {\doibase 10.1103/physreva.69.031602}
  {\bibfield  {journal} {\bibinfo  {journal} {Phys.~Rev.~A}\ }\textbf {\bibinfo
  {volume} {69}},\ \bibinfo {pages} {031602(R)} (\bibinfo {year}
  {2004})}\BibitemShut {NoStop}%
\bibitem [{\citenamefont {Grimm}(2008)}]{Grimm2008ufg}%
  \BibitemOpen
  \bibfield  {author} {\bibinfo {author} {\bibfnamefont {R.}~\bibnamefont
  {Grimm}},\ }\enquote {\bibinfo {title}
  {\href{https://arxiv.org/abs/cond-mat/0703091}{Ultracold Fermi gases in the
  BEC-BCS crossover: a review from the Innsbruck perspective}},}\ in\
  \href@noop {} {\emph {\bibinfo {booktitle} {Ultra-cold Fermi Gases}}},\
  \bibinfo {editor} {edited by\ \bibinfo {editor} {\bibfnamefont
  {M.}~\bibnamefont {Inguscio}}, \bibinfo {editor} {\bibfnamefont
  {W.}~\bibnamefont {Ketterle}}, \ and\ \bibinfo {editor} {\bibfnamefont
  {C.}~\bibnamefont {Salomon}}}\ (\bibinfo {year} {2008})\ \bibinfo {note}
  {{P}roceedings of the International School of Physics ``Enrico Fermi'',
  Course CLXIV, Varenna, 20-30 June 2006}\BibitemShut {NoStop}%
\bibitem [{\citenamefont {Petrov}(2003)}]{Petrov2003tbp}%
  \BibitemOpen
  \bibfield  {author} {\bibinfo {author} {\bibfnamefont {D.~S.}\ \bibnamefont
  {Petrov}},\ }\href
  {https://journals.aps.org/pra/abstract/10.1103/PhysRevA.67.010703} {\bibfield
   {journal} {\bibinfo  {journal} {Phys. Rev. A}\ }\textbf {\bibinfo {volume}
  {67}},\ \bibinfo {pages} {010703} (\bibinfo {year} {2003})}\BibitemShut
  {NoStop}%
\bibitem [{\citenamefont {Ketterle}\ and\ \citenamefont {{van
  Druten}}(1996)}]{Ketterle1996eco}%
  \BibitemOpen
  \bibfield  {author} {\bibinfo {author} {\bibfnamefont {W.}~\bibnamefont
  {Ketterle}}\ and\ \bibinfo {author} {\bibfnamefont {N.~J.}\ \bibnamefont
  {{van Druten}}},\ }\href {\doibase 10.1016/S1049-250X(08)60101-9} {\bibfield
  {journal} {\bibinfo  {journal} {Adv. At. Mol. Opt. Phys.}\ }\textbf {\bibinfo
  {volume} {37}},\ \bibinfo {pages} {181} (\bibinfo {year} {1996})}\BibitemShut
  {NoStop}%
\bibitem [{\citenamefont {Moerdijk}\ \emph {et~al.}(1995)\citenamefont
  {Moerdijk}, \citenamefont {Verhaar},\ and\ \citenamefont
  {Axelsson}}]{Moerdijk1995riu}%
  \BibitemOpen
  \bibfield  {author} {\bibinfo {author} {\bibfnamefont {A.~J.}\ \bibnamefont
  {Moerdijk}}, \bibinfo {author} {\bibfnamefont {B.~J.}\ \bibnamefont
  {Verhaar}}, \ and\ \bibinfo {author} {\bibfnamefont {A.}~\bibnamefont
  {Axelsson}},\ }\href
  {https://journals.aps.org/pra/abstract/10.1103/PhysRevA.51.4852} {\bibfield
  {journal} {\bibinfo  {journal} {Phys. Rev. A}\ }\textbf {\bibinfo {volume}
  {51}},\ \bibinfo {pages} {4852} (\bibinfo {year} {1995})}\BibitemShut
  {NoStop}%
\bibitem [{\citenamefont {Gribakin}\ and\ \citenamefont
  {Flambaum}(1993)}]{Gribakin1993cot}%
  \BibitemOpen
  \bibfield  {author} {\bibinfo {author} {\bibfnamefont {G.~F.}\ \bibnamefont
  {Gribakin}}\ and\ \bibinfo {author} {\bibfnamefont {V.~V.}\ \bibnamefont
  {Flambaum}},\ }\href
  {https://journals.aps.org/pra/abstract/10.1103/PhysRevA.48.546} {\bibfield
  {journal} {\bibinfo  {journal} {Phys. Rev. A}\ }\textbf {\bibinfo {volume}
  {48}},\ \bibinfo {pages} {546} (\bibinfo {year} {1993})}\BibitemShut
  {NoStop}%
\bibitem [{\citenamefont {Julienne}\ and\ \citenamefont
  {Hutson}(2014)}]{Julienne2014}%
  \BibitemOpen
  \bibfield  {author} {\bibinfo {author} {\bibfnamefont {P.~S.}\ \bibnamefont
  {Julienne}}\ and\ \bibinfo {author} {\bibfnamefont {J.~M.}\ \bibnamefont
  {Hutson}},\ }\href
  {https://journals.aps.org/pra/abstract/10.1103/PhysRevA.89.052715} {\bibfield
   {journal} {\bibinfo  {journal} {Phys. Rev. A}\ }\textbf {\bibinfo {volume}
  {89}} (\bibinfo {year} {2014})}\BibitemShut {NoStop}%
\bibitem [{\citenamefont {Bartenstein}\ \emph {et~al.}(2005)\citenamefont
  {Bartenstein}, \citenamefont {Altmeyer}, \citenamefont {Riedl}, \citenamefont
  {Geursen}, \citenamefont {Jochim}, \citenamefont {Chin}, \citenamefont
  {Denschlag}, \citenamefont {Grimm}, \citenamefont {Simoni}, \citenamefont
  {Tiesinga}, \citenamefont {Williams},\ and\ \citenamefont
  {Julienne}}]{Bartenstein:2005}%
  \BibitemOpen
  \bibfield  {author} {\bibinfo {author} {\bibfnamefont {M.}~\bibnamefont
  {Bartenstein}}, \bibinfo {author} {\bibfnamefont {A.}~\bibnamefont
  {Altmeyer}}, \bibinfo {author} {\bibfnamefont {S.}~\bibnamefont {Riedl}},
  \bibinfo {author} {\bibfnamefont {R.}~\bibnamefont {Geursen}}, \bibinfo
  {author} {\bibfnamefont {S.}~\bibnamefont {Jochim}}, \bibinfo {author}
  {\bibfnamefont {C.}~\bibnamefont {Chin}}, \bibinfo {author} {\bibfnamefont
  {J.~H.}\ \bibnamefont {Denschlag}}, \bibinfo {author} {\bibfnamefont
  {R.}~\bibnamefont {Grimm}}, \bibinfo {author} {\bibfnamefont
  {A.}~\bibnamefont {Simoni}}, \bibinfo {author} {\bibfnamefont
  {E.}~\bibnamefont {Tiesinga}}, \bibinfo {author} {\bibfnamefont {C.~J.}\
  \bibnamefont {Williams}}, \ and\ \bibinfo {author} {\bibfnamefont {P.~S.}\
  \bibnamefont {Julienne}},\ }\href {\doibase 10.1103/physrevlett.94.103201}
  {\bibfield  {journal} {\bibinfo  {journal} {Phys.~Rev.~Lett.}\ }\textbf
  {\bibinfo {volume} {94}},\ \bibinfo {pages} {103201} (\bibinfo {year}
  {2005})}\BibitemShut {NoStop}%
\bibitem [{\citenamefont {Chin}\ \emph
  {et~al.}(2010{\natexlab{b}})\citenamefont {Chin}, \citenamefont {Grimm},
  \citenamefont {Julienne},\ and\ \citenamefont {Tiesinga}}]{Chin:2010}%
  \BibitemOpen
  \bibfield  {author} {\bibinfo {author} {\bibfnamefont {C.}~\bibnamefont
  {Chin}}, \bibinfo {author} {\bibfnamefont {R.}~\bibnamefont {Grimm}},
  \bibinfo {author} {\bibfnamefont {P.}~\bibnamefont {Julienne}}, \ and\
  \bibinfo {author} {\bibfnamefont {E.}~\bibnamefont {Tiesinga}},\ }\href
  {\doibase 10.1103/revmodphys.82.1225} {\bibfield  {journal} {\bibinfo
  {journal} {Rev.~Mod.~Phys.}\ }\textbf {\bibinfo {volume} {82}},\ \bibinfo
  {pages} {1225} (\bibinfo {year} {2010}{\natexlab{b}})}\BibitemShut {NoStop}%
\end{thebibliography}

\end{document}